\documentclass[useAMS,usenatbib]{mn2e}
\usepackage{graphicx}
\title[Does obscured AGN fraction depend on luminosity?]{Does the
  obscured AGN fraction really depend on luminosity?}

\author[S. Sazonov et al.]{S. Sazonov$^{1}$\thanks{E-mail:
sazonov@iki.rssi.ru}, E. Churazov$^{2,1}$ and R. Krivonos$^{1}$\\
$^{1}$Space Research Institute, Russian Academy of Sciences,
Profsoyuznaya 84/32, 117997 Moscow, Russia\\
$^{2}$Max-Planck-Institut f\"ur Astrophysik,
Karl-Schwarzschild-Str. 1, 85740 Garching bei M\"unchen, Germany
}

\newcommand{\beq}{\begin{equation}}
\newcommand{\eeq}{\end{equation}}
\newcommand{\beqa}{\begin{eqnarray}}
\newcommand{\eeqa}{\end{eqnarray}}

\newcommand{\lesssim}{\hbox{\rlap{\hbox{\lower4pt\hbox{$\sim$}}}\hbox{$<$}}}
\newcommand{\gtrsim}{\mathrel{\hbox{\rlap{\hbox{\lower4pt\hbox{$\sim$}}}\hbox{$>$}}}} 

\newcommand{\nh}{N_{\rm{H}}}
\newcommand{\nhi}{N_{\rm{H},i}}
\newcommand{\vmax}{V_{\rm{max}}}
\newcommand{\vmaxi}{V_{\rm{max},i}}
\newcommand{\vmaxunobsc}{V_{\rm{max,unobsc}}}
\newcommand{\vmaxobsc}{V_{\rm{max,obsc}}}
\newcommand{\nheq}{N_{\rm{H,eq}}}
\newcommand{\nheqi}{N_{\rm{H,eq},i}}
\newcommand{\nhmin}{N_{\rm{H,min}}}
\newcommand{\nhmax}{N_{\rm{H,max}}}
\newcommand{\ecut}{E_{\rm{cut}}}

\newcommand{\fobs}{f_{\rm{obs}}}
\newcommand{\fdet}{f_{\rm{det}}}

\newcommand{\lobs}{L_{\rm{obs}}}
\newcommand{\lobsi}{L_{\rm{obs},i}}
\newcommand{\lintr}{L_{\rm{intr}}}
\newcommand{\lintri}{L_{\rm{intr},i}}
\newcommand{\lb}{L_\ast}

\newcommand{\robsc}{R_{\rm{obsc}}}
\newcommand{\runobsc}{R_{\rm{unobsc}}}

\newcommand{\nobsc}{N_{\rm{obsc}}}
\newcommand{\nunobsc}{N_{\rm{unobsc}}}

\begin{document}

\maketitle

\label{firstpage}

\begin{abstract}

We use a sample of 151 local non-blazar AGN selected from the \textit{INTEGRAL}
all-sky hard X-ray survey to investigate if the observed declining
trend of the fraction of obscured (i.e. showing X-ray absorption) AGN
with increasing luminosity is mostly an intrinsic or selection
effect. Using a torus-obscuration model, we demonstrate that in 
addition to negative bias, due to absorption in the torus, in finding
obscured AGN in hard X-ray flux limited surveys, there is also 
positive bias in finding unobscured AGN, due to Compton reflection
in the torus. These biases can be even stronger taking into account
plausible intrinsic collimation of hard X-ray emission along the axis
of the obscuring torus. Given the AGN luminosity function, which
steepens at high luminosities, these observational biases lead to a
decreasing observed fraction of obscured AGN with increasing
luminosity even if this fraction has no intrinsic luminosity
dependence. We find that if the central hard X-ray source in AGN is
isotropic, the intrinsic (i.e. corrected for biases) obscured AGN
fraction still shows a declining trend with luminosity, although the
intrinsic obscured fraction is significantly larger than the observed
one: the actual fraction is larger than $\sim 85$\% at $L\lesssim
10^{42.5}$~erg~s$^{-1}$ (17--60~keV), and decreases to $\lesssim 60$\%
at $L\gtrsim 10^{44}$~erg~s$^{-1}$. In terms of the half-opening angle
$\theta$ of an obscuring torus, this implies that $\theta\lesssim
30^\circ$ in lower-luminosity AGN, and $\theta\gtrsim 45^\circ$ in
higher-luminosity ones. If, however, the emission from the central
SMBH is collimated as $dL/d\Omega\propto\cos\alpha$, the intrinsic
dependence of the obscured AGN fraction is consistent with a
luminosity-independent torus half-opening angle $\theta\sim 30^\circ$. 

\end{abstract}


\section{Introduction}
\label{s:intro}

A lot of recent studies based on X-ray and hard X-ray extragalactic
surveys have demonstrated that the fraction of X-ray absorbed
(hereafter referred to as obscured) active galactic nuclei (AGN)
decreases with increasing observed X-ray luminosity, at least at
$\gtrsim 10^{42}$~erg~s$^{-1}$, both in the local ($z\approx 0$) and
high-redshift Universe
(\citealt{uedetal03,steetal03,hasetal04,sazrev04,lafetal05,sazetal07,hasetal08,becetal09,brinan11,buretal11,uedetal14,airetal15,bucetal15};
note also earlier evidence, \citealt{lawelv82}). This might indicate
that the opening angle of the (presumably) toroidal obscuring
structure -- the key element of AGN unification schemes -- increases
with AGN luminosity, for example due to feedback of the central
supermassive black hole (SMBH) on the accretion flow.

Could the observed luminosity dependence of the obscured AGN fraction
arise due to selection effects? This question has been occasionally
raised before (e.g. \citealt{maylaw13}) and is prompted by the fact
that even hard X-ray ($\gtrsim 10$~keV) surveys, which are usually
flux (or signal-to-noise ratio) limited, should be biased 
against detection of Compton-thick AGN, i.e. objects viewed through
absorption column density $\nh\gtrsim 10^{24}$~cm$^{-2}$, let alone
X-ray surveys at energies below 10~keV which must be biased
against even Compton-thin obscured sources. Due to this detection
bias, the observed fraction of obscured AGN is expected to be lower 
than the intrinsic fraction of such objects. Furthermore, this effect
may depend on luminosity, somehow reflecting the shape of the AGN
luminosity function (LF). In fact, as discussed by \cite{lawelv10},
some mid-infrared selected, radio selected and 
volume-limited AGN samples do \textit{not} demonstrate any clear
luminosity dependence of the proportion of type 1 (i.e. containing
broad emission lines in the optical spectrum) and type 2 AGN.

Although there have been previous attempts
\citep{uedetal03,maletal09,buretal11,uedetal14} to take into account
detection biases when estimating the space density of obscured AGN based
on hard X-ray surveys, they were, in our view, not fully
self-consistent and/or used too small samples of hard X-ray selected
AGN. It is our goal here to improve on both of these aspects. 

The purpose of the present study is to i) evaluate the impact on the
observed hard X-ray LF and observed luminosity dependence of the
obscured AGN fraction of the \textit{negative bias} for obscured AGN 
discussed above and a \textit{positive bias} that we demonstrate
likely exists for unobscured AGN, and ii) reconstruct the
\textit{intrinsic} dependence of the fraction of obscured AGN on
luminosity in the local Universe. Our treatment is based on a
realistic torus-like obscuration model and makes use of the
\textit{INTEGRAL}/IBIS 7-year (2002--2009) hard X-ray survey of the
extragalactic sky. Our sample consists of $\sim 150$ local ($z\lesssim
0.2$) Seyfert galaxies and is highly complete and reliable. Although
there are now significantly larger hard X-ray selected samples of
local AGN, based on additional observations by \textit{INTEGRAL}/IBIS
and especially by \textit{Swift}/BAT, they currently suffer from
significant incompleteness as concerns identification and absorption
column density information. Most importantly, our sample is large
enough to contain a significant number, 17, of heavily obscured
($\nh\ge 10^{24}$~cm$^{-2}$) AGN, for which we use as much as possible 
$\nh$ estimates based on high-quality hard X-ray spectral data, in
particular from the \textit{NuSTAR} observatory, which has recently
been systematically observing AGN discovered in the \textit{Swift}/BAT
and \textit{INTEGRAL}/IBIS hard X-ray surveys.

We adopt a $\Lambda CDM$ cosmological model with $\Omega_{\rm m}=0.3$
and $H_0=70$~km~s$^{-1}$~Mpc$^{-3}$.
 
\section{The INTEGRAL AGN sample}
\label{s:sample}

\begin{figure}
\centering
\includegraphics[width=\columnwidth,bb=0 180 580 700]{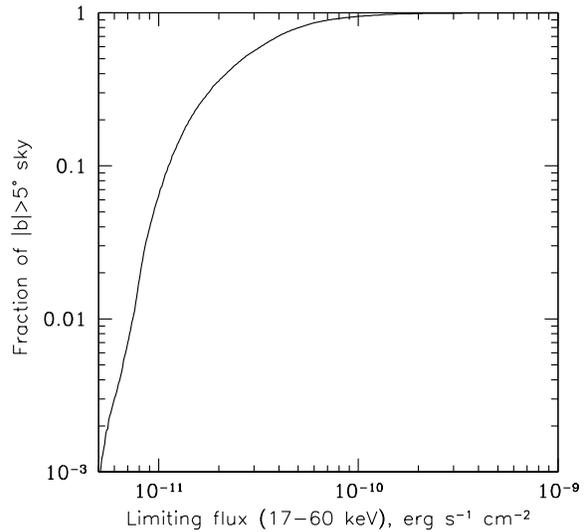}
\caption{Cumulative fraction of the extragalatic ($|b|>5^\circ$) sky 
as a function of flux limit in the \textit{INTEGRAL} 7-year survey. 
}
\label{fig:area_flux}
\end{figure}

We use the catalogue of sources \citep{krietal10b} from the
\textit{INTEGRAL}/IBIS 7-year all-sky hard X-ray survey (hereafer, the
\textit{INTEGRAL} 7-year survey, \citealt{krietal10a}). To minimise
possible biases in our study of the local AGN population due to
remaining unidentified \textit{INTEGRAL} sources and objects with
missing distance and/or X-ray absorption information, we exclude from
the consideration the Galactic plane region ($|b|<5^\circ$). The
catalogue is composed of sources detected on the time-averaged
(December 2002 -- July 2009) 17--60~keV map of the sky and is
significance limited ($5\sigma$). The corresponding flux limit varies
over the sky: $\fdet<2.6\times 10^{-11}$ ($<7\times
10^{-11}$~erg~s$^{-1}$~cm$^{-2}$) for 50\% (90\%) of the extragalactic
($|b|>5^\circ$) sky (see Fig.~\ref{fig:area_flux}).

The main properties of the \textit{INTEGRAL} 7-year survey and of the
corresponding catalogue of sources were described by
\citet{krietal10a,krietal10b}. Using this catalogue, \cite{sazetal10}
made preliminary estimates of the hard X-ray LF of local AGN and the
dependence of the obscured AGN fraction on luminosity. Subsequent
follow-up efforts by different teams have resulted in additional
identifications, classifications, distance measurements and X-ray
absorption column estimates for many \textit{INTEGRAL} sources, which
has significantly improved the quality of the catalogue, as detailed
below. 

\begin{figure}
\centering
\includegraphics[width=\columnwidth,bb=0 180 580 710]{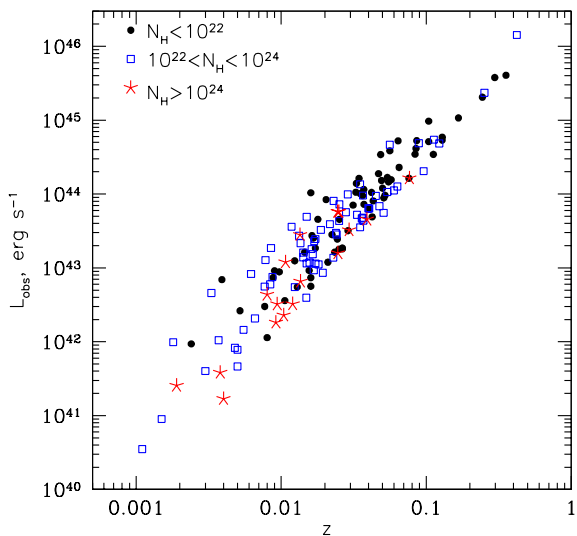}
\caption{Observed hard X-ray (17--60~keV) luminosity vs. redshift for
  non-blazar AGN from the \textit{INTEGRAL} 7-year survey. Filled
  circles, empty squares and stars denote unobscured, lightly obscured
  and heavily obscured objects, respectively.
}
\label{fig:z_lum}
\end{figure}

The final sample used here consists of 151 non-blazar
(i.e. Seyfert-like) AGN (see Table~\ref{tab:agn} in
Appendix~\ref{s:catalog}), with blazars (15 in total) being excluded
from the analysis. The sample is highly complete, as there are only 4 
sources at $|b|>5^\circ$ from the \textit{INTEGRAL} 7-year catalogue
that remain unidentified. Moreover, all of our AGN have known
distances and reliable estimates of their absorption columns based on
X-ray spectroscopy. As illustrated in Fig.~\ref{fig:z_lum}, our sample
is mostly local, with 146 out of the 151 objects being located at
$z<0.2$, and spans about 5 decades in (observed) luminosity, from
$\lobs\sim 10^{41}$ to $\sim 10^{46}$~erg~s$^{-1}$ (hereafter, all
luminosities are in the 17--60~keV energy band, unless specified
otherwise). 

We note that although we used the most up-to-date information from the
NASA/IPAC Extragalactic Database (NED) and recent literature to
remove blazars from our AGN sample, we cannot rule out that
some of our objects have blazar-like properties, i.e. their observed
hard X-ray emission contains a significant contribution from a
relativistic jet. The most suspicious in this respect are objects
classified as broad-line (i.e. presumably oriented 
towards us) radio galaxies. There are 6 such AGN in our sample:
3C~111, 3C~120, Pic~A, 3C~390.3, 4C~+74.26 and S5~2116+81. All of them
have $\lobs>10^{44}$~erg~s$^{-1}$ (but $<10^{45}$~erg~s$^{-1}$),
i.e. belong to the high-luminosity part of the sample. However, the
total number of objects with $\lobs>10^{44}$~erg~s$^{-1}$ is much larger:
42. This suggests that possible incomplete filtering of the sample
from blazars is unlikely to significantly affect the results and
conclusions of this work.

\subsection{Absorption columns, heavily obscured AGN}
\label{s:nh}

\begin{table*}
\caption{Heavily obscured AGN from the INTEGRAL 7-year survey
\label{tab:ctagn}
}

\begin{tabular}{l|r|l|c|l}
\hline
\multicolumn{1}{c}{Object} & 
\multicolumn{1}{c}{$D$} & 
\multicolumn{1}{c}{$L_{\rm obs}$} & 
\multicolumn{1}{c}{$\nh$} & 
\multicolumn{1}{c}{Reference for $\nh$} \\

 & \multicolumn{1}{c}{Mpc} & \multicolumn{1}{c}{erg~s$^{-1}$} & \multicolumn{1}{c}{cm$^{-2}$} & \\
\hline

SWIFT~J0025.8+6818 &  52.0 & $3.2\times 10^{42}$ &$> 10^{25}$ &
\textit{NuSTAR} \citep{krietal15} \\ 
NGC~1068           &  12.3 & $3.8\times 10^{41}$ &$> 10^{25}$ &
\textit{NuSTAR} \citep{bauetal14} \\
NGC~1194           &  59.0 & $6.6\times 10^{42}$ & $\sim 10^{24}$? &
\textit{XMM-Newton} (below 10 keV, \citealt{greetal08}) \\
CGCG~420-015       & 129.2 & $3.3\times 10^{43}$ &$> 10^{25}$ &
\textit{NuSTAR} \citep{krietal15} \\
MRK~3              &  58.6 & $2.8\times 10^{43}$ & $10^{24}$ &
\textit{Suzaku} \citep{ikeetal09} \\
IGR~J09253+6929    & 172.6 & $4.5\times 10^{43}$ & $>10^{24}$? &
low X-ray/hard X-ray flux ratio (\textit{Swift}/XRT+\textit{INTEGRAL}/IBIS) \\
NGC~3081           &  28.6 & $4.4\times 10^{42}$ & $10^{24}$ &
\textit{Suzaku} \citep{eguetal11} \\
NGC~3281           &  46.3 & $1.2\times 10^{43}$ & $2\times 10^{24}$ &
\textit{BeppoSAX} \citep{vigcom02} \\
ESO~506-G027       & 109.5 & $5.8\times 10^{43}$ & $10^{24}$ &
\textit{Suzaku} \citep{winetal09b} \\
NGC~4939           &  34.7 & $2.3\times 10^{42}$ &$>10^{25}$? &
\textit{BeppoSAX} \citep{maietal98}, but varied to $\nh=1.5\times
10^{23}$~cm$^{-2}$ \\ 
 & & & & (\textit{XMM-Newton}, below 10~keV, \citealt{guaetal05a})\\
NGC~4945           &   3.4 & $2.6\times 10^{41}$ & $4\times 10^{24}$ &
\textit{NuSTAR} \citep{pucetal14,brietal15}, \textit{Suzaku}
\citep{yaqetal12} \\ 
IGR~J14175$-$4641  & 348.3 & $1.6\times 10^{44}$ &$>10^{24}$? &
low X-ray/hard X-ray flux ratio (\textit{Swift}/XRT+\textit{INTEGRAL}/IBIS) \\
NGC~5643           &  11.8 & $1.7\times 10^{41}$ & $>10^{25}$ &
\textit{NuSTAR} \citep{krietal15} \\
NGC~5728           &  24.8 & $3.2\times 10^{42}$ & $2\times 10^{24}$ &
\textit{NuSTAR} \citep{krietal15} \\
IGR~J14561$-$3738  & 107.7 & $1.6\times 10^{43}$ & $\sim 10^{24}$ &
\textit{Chandra}+\textit{INTEGRAL}/IBIS \citep{sazetal08} \\
ESO~137-G034       &  33.0 & $1.8\times 10^{42}$ & $>10^{25}$ &
\textit{Suzaku} \citep{cometal10} \\
NGC~6240           & 107.3 & $5.8\times 10^{43}$ & $2.5\times 10^{24}$ &
\textit{NuSTAR} \citep{krietal15} \\

\hline
\end{tabular}

\end{table*}

For the purposes of this study it is important to have maximally 
complete and reliable information on the X-ray absorption columns,
$\nh$, of the studied AGN. Our starting source of such information is
our previous papers on the \textit{INTEGRAL}/IBIS survey
\citep{sazetal07,sazetal12} as well as on the \textit{RXTE}
(3--20~keV) slew survey \citep{sazrev04}, but we have updated the
$\nh$ estimates in all cases where it was necessary and possible (see
Table~\ref{tab:agn}). 

For unobscured and lightly obscured ($\nh<10^{24}$~cm$^{-2}$)
sources, X-ray spectroscopy at energies below 10~keV is usually
sufficient for evaluating $\nh$. Such data do exist for all of our
sources and in most cases there are reliable published $\nh$ values,
which we adopt. Furthermore, if the absorption column is less 
than $10^{22}$~cm$^{-2}$, we adopt $\nh=0$ and consider such sources
unobscured.  

However, absorption column estimates based on X-ray data below 10~keV
become unreliable for strongly absorbed sources, having
$\nh\ge 10^{24}$~cm$^{-2}$. In such cases, we prefer to use results
from hard X-ray (above 10~keV) spectroscopy, whenever
possible. Specifically, our preference list of instruments is headed
by \textit{NuSTAR} -- the unique focusing hard X-ray telescope,
followed by Suzaku and then by all other currently operating or
previously flown hard X-ray missions. 

For five of the heavily obscured ($\nh\ge 10^{24}$~cm$^{-2}$) objects
and candidates, we carried out our own analysis of publicly available
\textit{NuSTAR} data \citep{krietal15}. Specifically, we fitted the
spectra by a sum of a strongly absorbed power-law component (with a
high-energy cutoff) and a disk-reflection continuum modelled with
\textit{pexrav} in XSPEC. The \textit{NuSTAR} spectra of
SWIFT~J0025.8+6818, CGCG~420-015 and NGC~5643 are consistent with
being fully reflection dominated (i.e. dominated by Compton-scattered
continuum), and so we prescribed $\nh>10^{25}$~cm$^{-2}$ to them. The
other two objects, NGC~5728 and NGC~6240, along with strong reflection
demonstrate a significant contribution from the primary component
suppressed by intrinsic absorption at the level of $\nh\sim
2$--$2.5\times 10^{24}$~cm$^{-2}$. More physically motivated AGN torus
models confirmed this qualitative result (see \citealt{krietal15} for
details). Our derived spectral parameters for CGCG~420-015, NGC~5643,
NGC~5728 and NGC~6240 are consistent with pre-\textit{NuSTAR}
estimates for these objects (\citealt{sevetal11}, \citealt{matetal13},
\citealt{cometal10}, \citealt{vigetal99}, respectively). 

In total, our sample consists of 67 unobscured
($\nh<10^{22}$~cm$^{-2}$) and 84 obscured ($\nh\ge 10^{22}$~cm$^{-2}$) AGN,
including 17 heavily obscured ($\nh\ge 10^{24}$~cm$^{-2}$) ones.  

Table~\ref{tab:ctagn} provides key information about our heavily
obscured AGN. For 7 of these, there are reliable $\nh$ estimates or
evidence that the source's spectrum is reflection-dominated (in which
case we adopt that $\nh> 10^{25}$~cm$^{-2}$) from \textit{NuSTAR}
observations. All but one (IGR~J14561$-$3738) of the remaining 10
objects are either planned to be observed by \textit{NuSTAR} soon or
have already been observed by this telescope but the data are
proprietary at the time of writing. However, for most of these sources
there exists fairly reliable information from other hard X-ray
missions indicating that $\nh\ge 10^{24}$~cm$^{-2}$ -- see
Table~\ref{tab:ctagn}.

Three of the objects included in our sample of heavily obscured AGN
are currently candidates rather than firmly established
representatives of this class: the quoted value $\nh\sim
10^{24}$~cm$^{-2}$ for NGC~1194 comes from X-ray data below 10~keV, 
whereas the presence of $\nh> 10^{24}$~cm$^{-2}$ absorption
columns in IGR~J09253+6929 and IGR~J14175$-$4641 is strongly suggested
by very low ($\lesssim 0.01$) X-ray/hard X-ray flux ratios that we
find for them from \textit{Swift}/XRT and \textit{INTEGRAL}/IBIS
data. Note that we initially used the same argument to regard another
source from this sample, SWIFT~J0025.8+6818, as a likely heavily
obscured AGN, and it indeed proved to be such once we analysed
\textit{NuSTAR} data. In the analysis below, we assume that
$\nh=3\times 10^{24}$~cm$^{-2}$ for both IGR~J09253+6929 and
IGR~J14175$-$4641. 

The most difficult case is that of NGC~4939, which manifested itself
as a reflection-dominated source ($\nh>10^{25}$~cm$^{-2}$) during
\textit{BeppoSAX} observations in 1997 \citep{maietal98}, but was
found to be in a Compton-thin state, with $\nh\sim 1.5\times
10^{23}$~cm$^{-2}$, by \textit{XMM-Newton} in 2001
\citep{guaetal05a}. We nevertheless treat NGC~4939 as a
reflection-dominated source in our analysis, in part because the hard
X-ray flux measured by \textit{INTEGRAL} for this source is similar to
that measured by \textit{BeppoSAX} but lower than the flux inferred 
from the \textit{XMM-Newton} observation and so
\textit{INTEGRAL} may have caught the source in a state similar to
that revealed by \textit{BeppoSAX}. Generally, we adopt
$\nh=10^{25}$~cm$^{-2}$ for reflection-dominated sources (there are in
total 6 such objects) in our analysis, although in reality the column
density in such objects may be even higher, say $\nh\sim
10^{26}$~cm$^{-2}$. 

We have thus obtained a fairly large and high-quality (in terms of
information on intrinsic obscuration) sample of heavily obscured
AGN. The high completeness and reliability of this sample are crucial
for our analysis below.

\section{Observed properties of local AGN}
\label{s:obs_properties}

We first consider a number of \textit{observed} properties of the
local AGN population using our \textit{INTEGRAL} sample. 

\begin{figure}
\centering
\includegraphics[width=\columnwidth,bb=0 180 580 710]{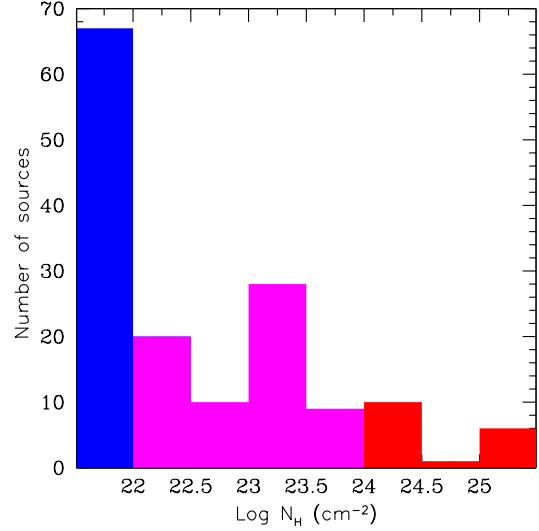}
\caption{Observed distribution of X-ray absorption columns for the
  \textit{INTEGRAL} AGN. Unobscured ($\nh<10^{22}$~cm$^{-2}$), lightly
  obscured ($10^{22}\le\nh<10^{24}$~cm$^{-2}$) and heavily obscured
  ($\nh\ge 10^{24}$~cm$^{-2}$) objects are shown in blue, magenta and
  red, respectively.
}
\label{fig:obs_nh_distr}
\end{figure}

\begin{figure}
\centering
\includegraphics[width=\columnwidth,bb=0 150 590 730]{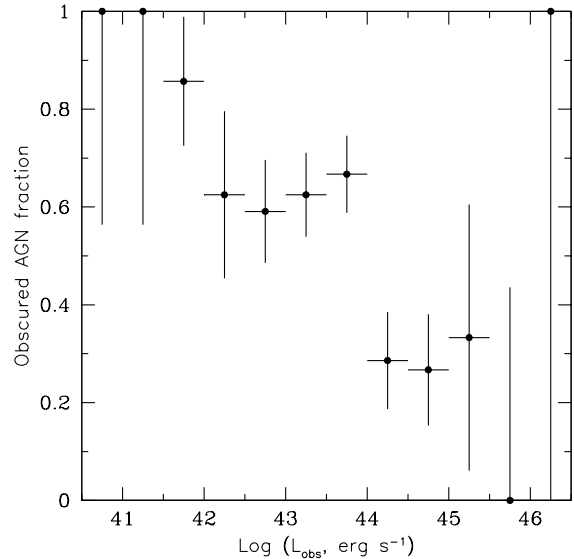}
\caption{Observed fraction of obscured ($\nh\ge 10^{22}$~cm$^{-2}$) AGN
  as a function of observed hard X-ray luminosity for the
  \textit{INTEGRAL} 7-year survey.
}
\label{fig:obs_obsc_frac}
\end{figure}

Fig.~\ref{fig:obs_nh_distr} shows the observed distribution of
absorption columns for our objects, while Fig.~\ref{fig:obs_obsc_frac} 
shows the observed dependence of the obscured AGN fraction on hard
X-ray luminosity. The latter was obtained by counting obscured and 
unobscured sources within specified luminosity bins and dividing the
first number by the sum of the two. One can clearly see a declining
trend of the obscured AGN fraction with increasing luminosity, which
is well known from previous studies.

\begin{figure}
\centering
\includegraphics[width=\columnwidth,bb=0 180 580 710]{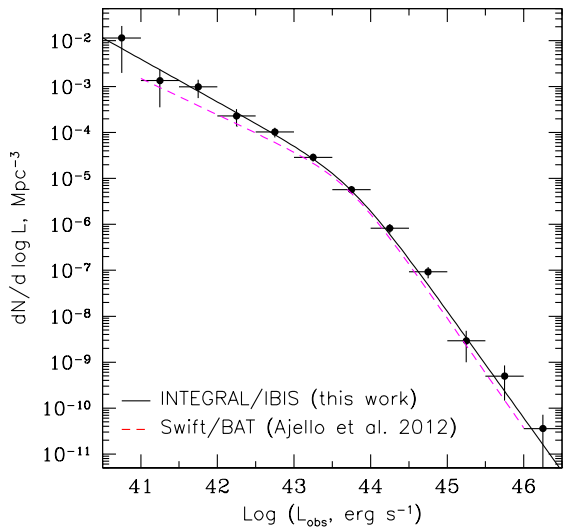}
\caption{Observed (in the \textit{INTEGRAL} 7-year survey) hard X-ray
  luminosity function of local AGN (filled circles) fitted by a broken
  power law (black solid line). The best-fit parameters are given in
  Table~\ref{tab:lumfunc}. For comparison the LF based on the
  \textit{Swift}/BAT survey \citep{ajeetal12} is shown by the magenta
  dashed line.  
} 
\label{fig:obs_lumfunc}
\end{figure}

We next calculated the observed hard X-ray LF, $\phi(\lobs)$ (number
of objects per Mpc$^{_3}$ per $\log\lobs$), of local AGN: both in  
binned and analytic form (see Fig.~\ref{fig:obs_lumfunc}). The
analytic LF model used throughout this study is a broken power law:
\beq
\frac{dN_{\rm AGN}}{d\log
  L}=\frac{A}{(L/\lb)^{\gamma_1}+(L/\lb)^{\gamma_2}}.
\label{eq:broken_pl}
\eeq
The binned LF was constructed using the standard $1/\vmax$ method,
whereas the best-fit values (and their uncertainties) of the
characteristic luminosity, $\lb$, and of the two slopes, $\gamma_1$
and $\gamma_2$, of the analytic model (see Table~\ref{tab:lumfunc})
were found using a maximum likelihood estimator (similarly to
\citealt{sazetal07}):
\beq
\mathcal{L}=-2\sum_{i}\ln\frac{\phi(\lobsi)\vmax(\lobsi)}
{\int\phi(\lobs)\vmax(\lobs) d\log\lobs},
\label{eq:like_lum}
\eeq
where $\lobsi$ are the observed luminosities of AGN in our sample, and
$\vmax(\lobs)$ is the volume of the Universe probed by the
\textit{INTEGRAL} 7-year survey for a given $\lobs$, which can be
calculated from the sky coverage curve (see Fig.~\ref{fig:area_flux}). 
The normalization of the analytic model is derived from the actual
number of objects in the sample.

Comparing this newly determined observed hard X-ray LF with our old
result \citep{sazetal07} based on a smaller set (66 vs. 151 objects)
of AGN detected with \textit{INTEGRAL}, we find good agreement between
the two, but the constraints on the LF parameters have now significantly
improved. We can also compare the \textit{INTEGRAL} LF with that
derived from a still larger (361~objects) sample of (mostly) local AGN
found in nearly the same energy band (15--55~keV) in the
\textit{Swift}/BAT survey \citep{ajeetal12}. As can be seen in
Fig.~\ref{fig:obs_lumfunc}, the two LFs are in good agreement with
each other.  
 
\begin{figure}
\centering
\includegraphics[width=\columnwidth,bb=0 170 560 680]{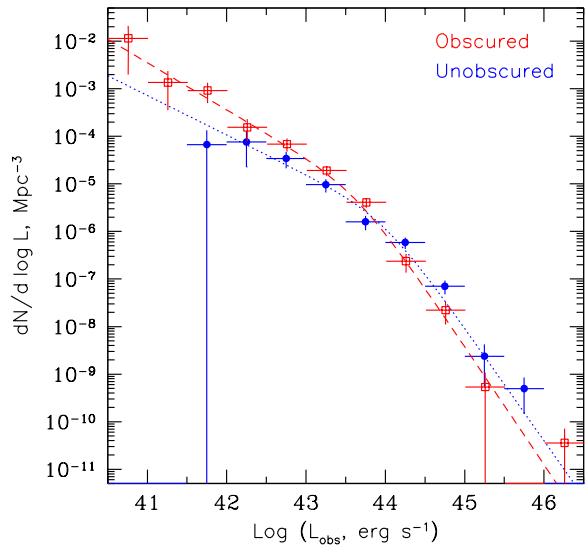}
\caption{Observed hard X-ray luminosity functions of unobscured
  ($\nh<10^{22}$~cm$^{-2}$, blue filled circles) and obscured ($\nh\ge
  10^{22}$~cm$^{-2}$, red empty squares) AGN, fitted by broken power
  laws (blue dotted and red dashed lines, respectively). The best-fit
  parameters are given in Table~\ref{tab:lumfunc}.
}
\label{fig:obs_lumfunc_nh}
\end{figure}

Finally, we calculated separately the observed LFs of unobscured and
obscured AGN (see Fig.~\ref{fig:obs_lumfunc_nh}). It can be seen that
these LFs are different in shape, as is verified by the best-fit
parameters of the corresponding analytic fits (see
Table~\ref{tab:lumfunc}).
 
\begin{table*}
\caption{Fits of different hard X-ray luminosity functions by a broken
  power law 
\label{tab:lumfunc}
}

\begin{tabular}{l|r|c|c|c|c|c|c}
\hline
\multicolumn{1}{c}{AGN} & \multicolumn{1}{c}{$N_{\rm AGN}$} &
\multicolumn{1}{c}{$\log\lb$} & 
\multicolumn{1}{c}{$\gamma_1$} & 
\multicolumn{1}{c}{$\gamma_2$} & 
\multicolumn{1}{c}{$A^{\rm a}$,} & 
\multicolumn{1}{c}{Num. density} &
\multicolumn{1}{c}{Lum. density} \\

\multicolumn{1}{c}{class} & & & & &
\multicolumn{1}{c}{$10^{-5}$~Mpc$^{-3}$} &
\multicolumn{1}{c}{($\log L=40.5$--$46.5$)} & 
\multicolumn{1}{c}{($\log L=40.5$--$46.5$)} \\

& & & & & & \multicolumn{1}{c}{$10^{-4}$~Mpc$^{-3}$} &
\multicolumn{1}{c}{$10^{39}$~erg~s$^{-1}$~Mpc$^{-3}$} \\

\hline
\multicolumn{8}{c}{Observed LF}\\
 
All & 151 & $43.74\pm0.19$ & $0.93\pm0.10$ & $2.33\pm0.15$ &
1.122 & 54 ($41\div 82$) & $1.57\pm0.20$ \\ 

Unobscured & 67 & $43.98\pm0.32$ & $0.83\pm0.18$ & $2.37\pm0.28$ &
0.243 & 10 ($7\div25$) & $0.45\pm0.08$ \\ 

Obscured & 84 & $43.65\pm0.21$ & $0.99\pm0.21$ & $2.48\pm0.22$ &
0.839 & 48 ($37\div 78$) & $1.14\pm0.18$ \\ 

\multicolumn{8}{c}{Intrinsic LF, isotropic emission, $\theta=30^\circ$}\\
 
All & 151 & $43.69\pm0.18$ & $0.89\pm0.10$ & $2.36\pm0.16$ &
1.806 & 61 ($47\div96$) & $1.97\pm0.25$ \\ 

Unobscured & 67 & $43.87\pm0.31$ & $0.84\pm0.18$ & $2.39\pm0.28$ &
0.201 & 7.0 ($5.1\div16.7$) & $0.30\pm0.05$ \\ 

Obscured & 84 & $43.62\pm0.21$ & $0.93\pm0.12$ & $2.45\pm0.21$ &
1.581 & 59 ($46\div101$) & $1.63\pm0.26$ \\ 

\multicolumn{8}{c}{Intrinsic LF, isotropic emission, $\theta=45^\circ$}\\

All & 151 & $43.71\pm0.19$ & $0.89\pm0.10$ & $2.36\pm0.16$ &
1.494 & 52 ($41\div82$) & $1.71\pm0.21$ \\ 

Unobscured & 67 & $43.90\pm0.31$ & $0.84\pm0.18$ & $2.39\pm0.27$ &
0.203 & 7.5 ($5.4\div 18.2$) & $0.32\pm0.06$ \\ 

Obscured & 84 & $43.62\pm0.21$ & $0.92\pm0.12$ & $2.44\pm0.21$ &
1.698 & 59 ($46\div105$) & $1.70\pm0.26$ \\ 

\multicolumn{8}{c}{Intrinsic LF, cosine-law emission, $\theta=30^\circ$}\\
 
All & 151 & $43.69\pm0.17$ & $0.90\pm0.10$ & $2.43\pm0.17$ &
1.593 & 57 ($44\div87$) & $1.79\pm0.21$ \\ 

Unobscured & 67 & $43.70\pm0.30$ & $0.86\pm0.17$ & $2.42\pm0.28$ &
0.188 & 5.3 ($3.9\div 11.6$) & $0.19\pm0.03$ \\ 

Obscured & 84 & $43.68\pm0.21$ & $0.93\pm0.12$ & $2.45\pm0.21$ &
1.565 & 66 ($51\div114$) & $1.88\pm0.30$ \\ 

\hline
\end{tabular}

$^{\rm a}$ The normalization $A$ is given without an error because
this parameter is strongly correlated with the others.

\end{table*}

\section{Intrinsic properties of local AGN}
\label{s:intr_properties}

The observed LF just discussed has not been corrected for any effects
associated with absorption or scattering of hard X-rays emitted by the
AGN central source on the way between the source and the observer, nor
for any intrinsic anisotropy of the emission generated by the central
source in AGN. This observed LF is expected to be affected by
absorption bias: an obscured AGN will be inferred to have a lower
luminosity, $\lobs=\fobs\times 4\pi D^2$ (here, $\fobs$ is the
measured hard X-ray flux and $D$ is the distance to the source), than
its intrinsic (i.e. emitted by the central source) luminosity,
$\lintr$, and a source like this can be found in a flux-limited hard
X-ray survey within a smaller volume of the Universe than it would be
in the absence of X-ray absorption: $\vmax(\lobs)/\vmax(\lintr)\approx
(\lobs/\lintr)^{3/2}$. Here, the approximation simbol reflects the
fact that AGN obscuration may also affect the shape of the measured
X-ray spectrum and thus the number of photons recorded by a given
detector with its specific energy response. On the other hand, as
discussed below, unobscured AGN are expected to have higher observed
luminosities than their intrinsic angular-averaged luminosities and
can thus be detected within a larger $\vmax$. We can correct for both
of these biases and obtain an intrinsic hard X-ray LF of local AGN. To
this end, we use a physically motivated obscuration model described
below. 

\subsection{Torus model and AGN spectra}
\label{s:torus}

\begin{figure}
\centering
\includegraphics[width=\columnwidth,bb=100 320 550 580]{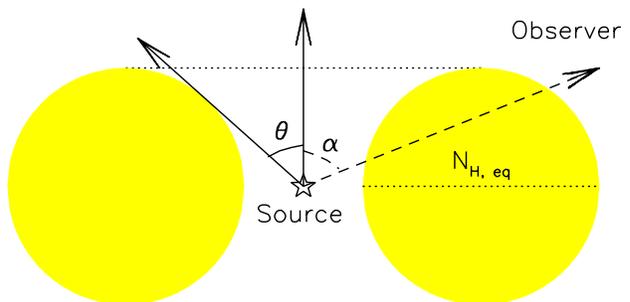}
\caption{Torus model.
} 
\label{fig:torus}
\end{figure}

We have a developed a Monte-Carlo code for modelling AGN X-ray spectra
modified by reprocessing in a toroidal structure of gas. The adopted
geometry (see Fig.~\ref{fig:torus}) is similar to that used in other
existing models, e.g. \cite{ikeetal09,muryaq09,brinan11}. The key
assumptions of our model are: 
\begin{enumerate}

\item
The geometrical shape is that of a ring torus;

\item
The gas is homogeneous, cold, neutral and of normal cosmic
chemical composition;

\item
The X-ray spectrum emitted by the central source is a power law with
an exponential cutoff, $dN/dE\propto E^{-\Gamma} e^{-E/\ecut}$, with
$\Gamma=1.8$ and $\ecut=200$~keV;

\item
The central (point-like) source is either isotropic,
$d\lintr/d\Omega={\rm const}$ -- hereafter, Model~A, or emitting
according to Lambert's law, $d\lintr/d\Omega\propto\cos\alpha$, where
$\alpha$ is the viewing angle with respect to the axis of the torus --
hereafter, Model~B.

\end{enumerate}

The introduction of Model~B is an important aspect of the present
study and is motivated by the common belief that the hard X-ray
emission observed from AGN is produced by Comptonization of softer
emission from an accretion disk around a SMBH in a hot corona lying
above the disk. If such a corona has quasi-planar geometry, the hard
X-ray flux it produces will be collimated along the axis of
the disk/corona roughly as $F\propto\mu$ (the exact law being
dependent on the photon energy and the optical depth of the
corona, \citealt{pozetal83,suntit85}), where $\mu$ is the cosine of
the angle between the outgoing direction and the axis of the
disk/corona. Because the obscuring torus in turn is likely coaligned with
the accretion disk, the emergent hard X-ray radiation will be
collimated along the axis of the torus. In reality, a significant
fraction of the coronal emission is reflected by the underlying
accretion disk, but this also occurs preferentially along the axis of
the disk/torus \citep{magzdz95}. Since there is still significant 
uncertainty in the overall physical picture, we introduce a simple, energy
independent collimation factor $d\lintr/d\Omega\propto\cos\alpha$ to get an
idea of how strongly intrinsic collimation of hard X-ray emission can
affect observed properties of local AGN.
 
\begin{figure}
\centering
\includegraphics[width=\columnwidth,bb=0 170 590 730]{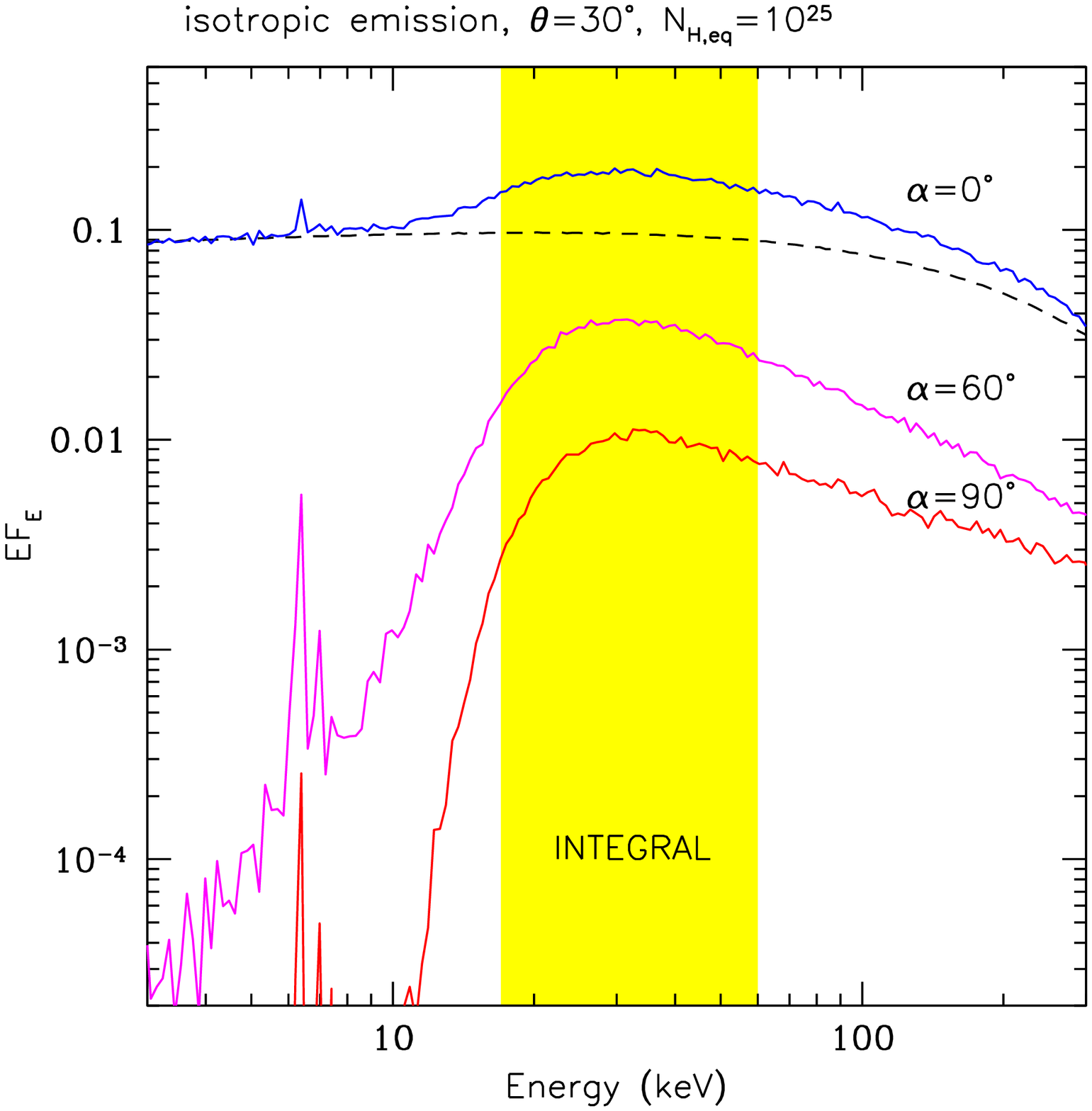}
\includegraphics[width=\columnwidth,bb=0 170 590 730]{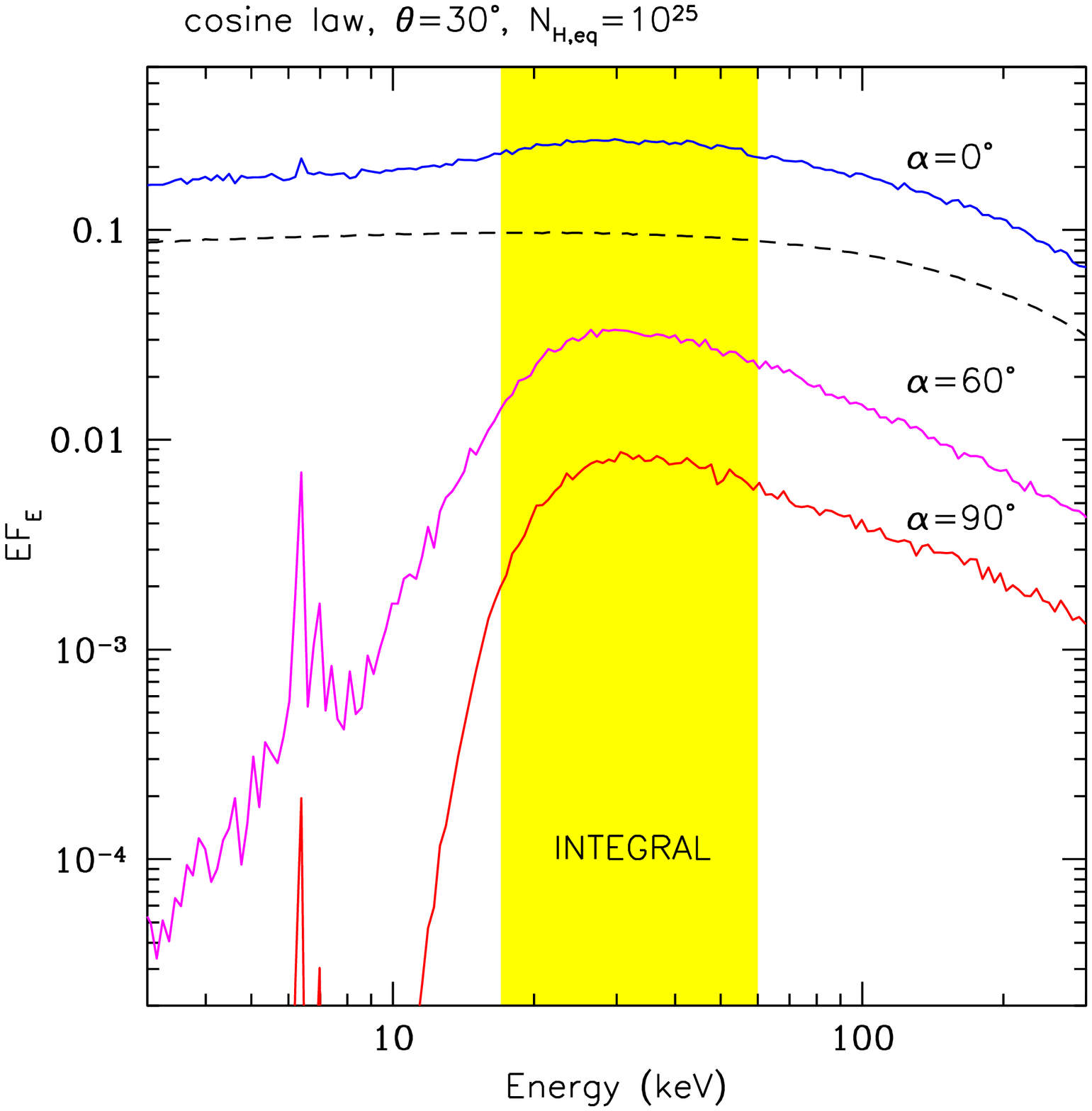}
\caption{\textit{Top:} Examples of simulated AGN spectra for Model~A,
  half-opening torus angle $\theta=30^\circ$ and equatorial column
  density $\nheq=10^{25}$~cm$^{-2}$, for various viewing angles. The
  dashed curve shows the intrinsic (angular-averaged) spectrum. The
  shaded area indicates the \textit{INTEGRAL}/IBIS energy band used
  for AGN selection in this work. \textit{Bottom:} The same, but for
  Model~B. 
}
\label{fig:spectra}
\end{figure}

Apart from the two alternatives for the angular dependence of
intrinsic emission (Model~A or Model~B), our model has three free
parameters: (i) the equatorial column density, $\nheq$ (the total
number of H atoms per cm$^{2}$ along an equatorial line of sight
between the central source and the observer, (ii) half-opening angle
of the torus, $\theta$, and (iii) the viewing angle relative to the
axis of the torus, $\alpha$ (see Fig.~\ref{fig:torus}). 

X-ray photons emitted by the central source can scatter multiple times
within the torus before they either get photoabsorbed in the gas or
escape from the system. Our radiative transfer calculations are based
on a method developed by \citet{chuetal08}. The gas
in the torus is assumed to be neutral, with the relative abundances of
all elements as in the Solar photosphere. The following processes are 
included in the simulations: photoelectric absorption, Rayleigh and
Compton scattering and fluorescence. Photoelectric absorption is
calculated using the data and approximations of \citet{veryak95} and 
\citet{veretal96}. For fluorescence we use the energies and yields from
\citet{kaamew93}. Compton and Rayleigh scattering are modelled using
differential cross sections provided by the GLECS package
\citep{kippen04} of the GEANT code \citep{agoetal03}. Namely, the 
Livermore Evaluated Photon Data Library (EPDL, see
\citealt{culetal90}) and the Klein-Nishina formula for free electrons
are used to calculate total cross-sections and the angular
distribution of scattered photons for each element.

Figure~\ref{fig:spectra} shows examples of emergent AGN spectra
simulated using our model of the obscuring torus. As expected, for
obscured AGN ($\alpha>\theta$), the observed hard X-ray flux can be
strongly attenuated relative to the emitted flux and for high absorption
columns ($\nh\gg 10^{24}$~cm$^{-2}$) the spectrum can become 
reflection-dominated, as the observer will mostly see emission
reflected from the inner walls of the torus rather than emission from
the central source transmitted through the torus. In this last
  case, there also appear strong iron K$\alpha$ and K$\beta$ fluorescent
  lines at 6.4~keV and 7.06~keV. These spectral properties and trends
for heavily obscured AGN are of course well known. 

We further see from Fig.~\ref{fig:spectra} that the spectra observed
from directions $\alpha<\theta$, corresponding to unobscured AGN, also  
differ from the intrinsic spectrum. Namely, they have an excess due to
Compton reflection of hard X-rays from the torus in the direction of
the observer. This hump is located approximately within the energy
band of 17--60~keV that we use for detecting AGN in the
\textit{INTEGRAL} survey. It is obvious that this Compton reflection
component should bias observed luminosities of unobscured AGN higher
in this and similar (e.g. \textit{Swift}/BAT) hard X-ray surveys. Any
intrinsic collimation of emission along the axis of the obscuring
torus will make this positive bias even stronger (see the spectrum for
Model~B and $\alpha\approx 0$ in the lower panel of
Fig.~\ref{fig:spectra}). This important aspect is frequently
overlooked in AGN population studies, even though a reflection
component is well known to be present in the hard X-ray spectra of
unobscured AGN.

\subsection{AGN detection bias}
\label{s:bias}

\begin{figure}
\centering
\includegraphics[width=\columnwidth,bb=0 180 580 730]{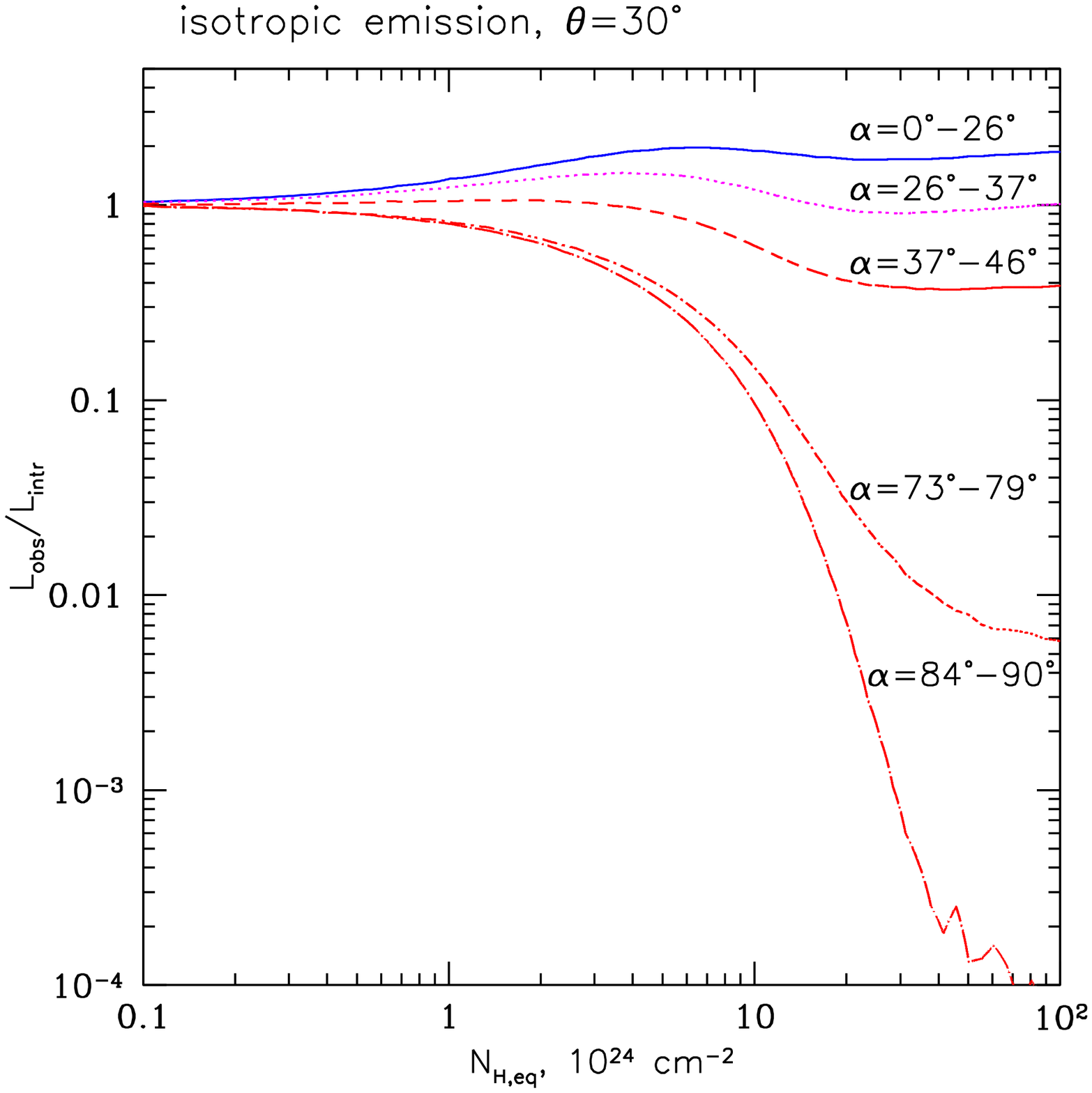}
\includegraphics[width=\columnwidth,bb=0 180 580 730]{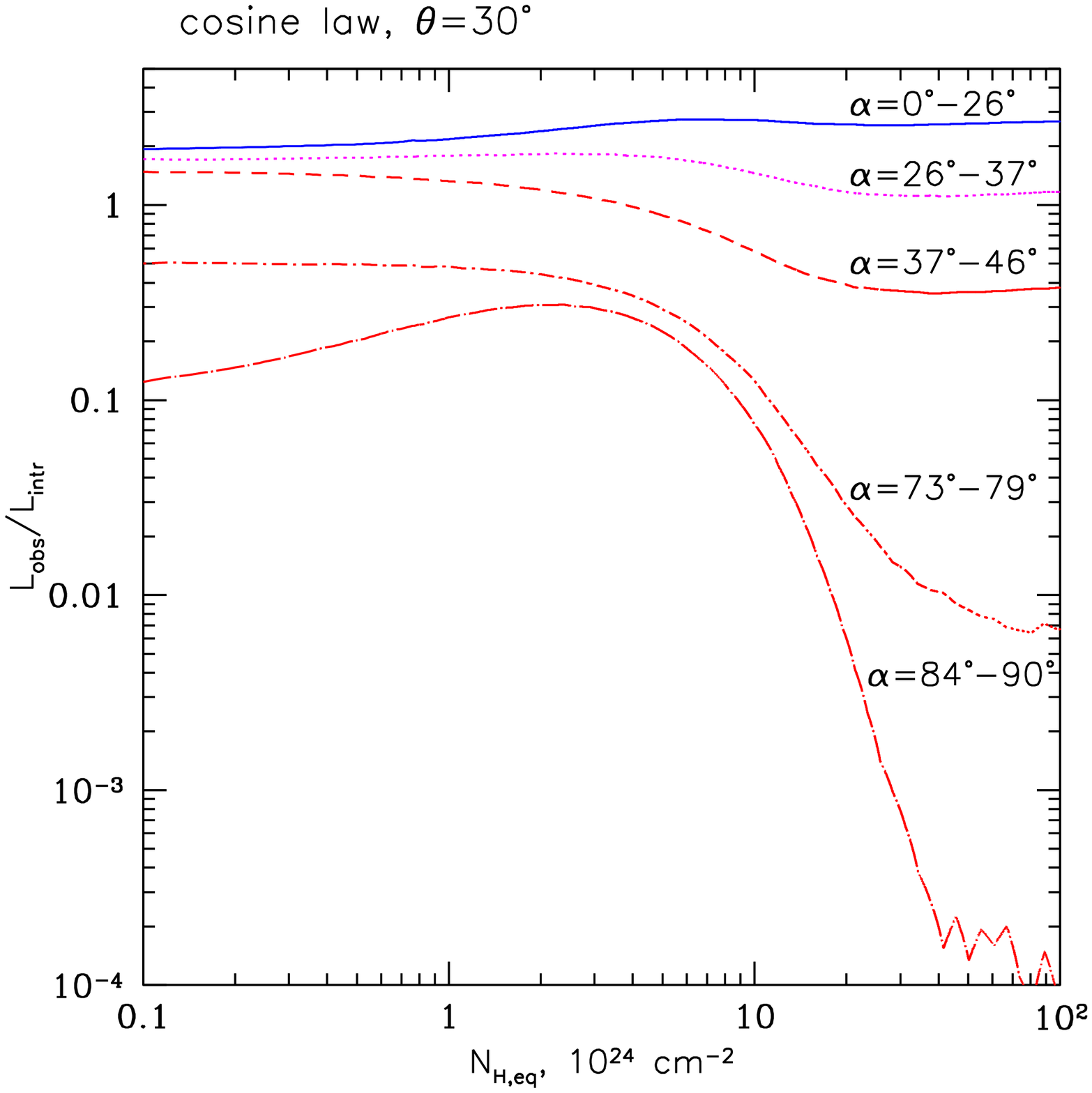}
\caption{\textit{Top:} Calculated ratio of observed to intrinsic
  (angular-averaged) luminosity in the 17--60~keV energy 
  band for a torus half-opening angle $\theta=30^\circ$, for different
  viewing angles ($\alpha$), as a function of the torus column density
  for Model~A. \textit{Bottom:} The same, but for Model~B.
}
\label{fig:flux_nh_angles}
\end{figure}

To quantify biases affecting detection of unobscured and obscured AGN
in the \textit{INTEGRAL} survey, we show in
Fig.~\ref{fig:flux_nh_angles}, for Model~A and Model~B, the ratio,
$R(\nheq,\theta,\alpha)=\lobs/\lintr$, of the observed to intrinsic
luminosity in the 17--60~keV energy band as a function of $\nheq$, for
a torus half-opening angle $\theta=30^\circ$ and several narrow ranges
of the viewing angle $\alpha$. One can see that $R$ is always larger
than unity, i.e. $\lobs>\lintr$, for unobscured AGN. For example, for
Model~A and $\alpha\approx 0$, $R$ increases from 1 to $\sim 2$ as
$\nheq$ increases to a few $10^{24}$~cm$^{-2}$ and remains at
approximately this level thereafter. This trend can be easily
understood: the amplitude of Compton reflection is expected to be
proportional to the torus optical depth, $\tau$, in the optically thin
regime ($\tau\ll 1$) and constant in the opposite case ($\tau\gg
1$). As regards obscured AGN ($\alpha>\theta$), $R$ decreases with
increasing $\nheq$ and increasing $\alpha$ (apart from a local maximum
at $\nheq\sim 3\times 10^{24}$~cm$^{-2}$ at near-equatorial directions
for Model~B -- due to the reflected component), as could be expected
due to the increasing attenuation of the transmitted component. The
most obvious and important difference of Model~B with respect to
Model~A is that the observed hard X-ray flux is anisotropic even in
the absence of an obscuring torus (i.e. for $\nh=0$) -- just due to
the initial collimation of emission. 

\begin{figure}
\centering
\includegraphics[width=\columnwidth,bb=0 150 580 710]{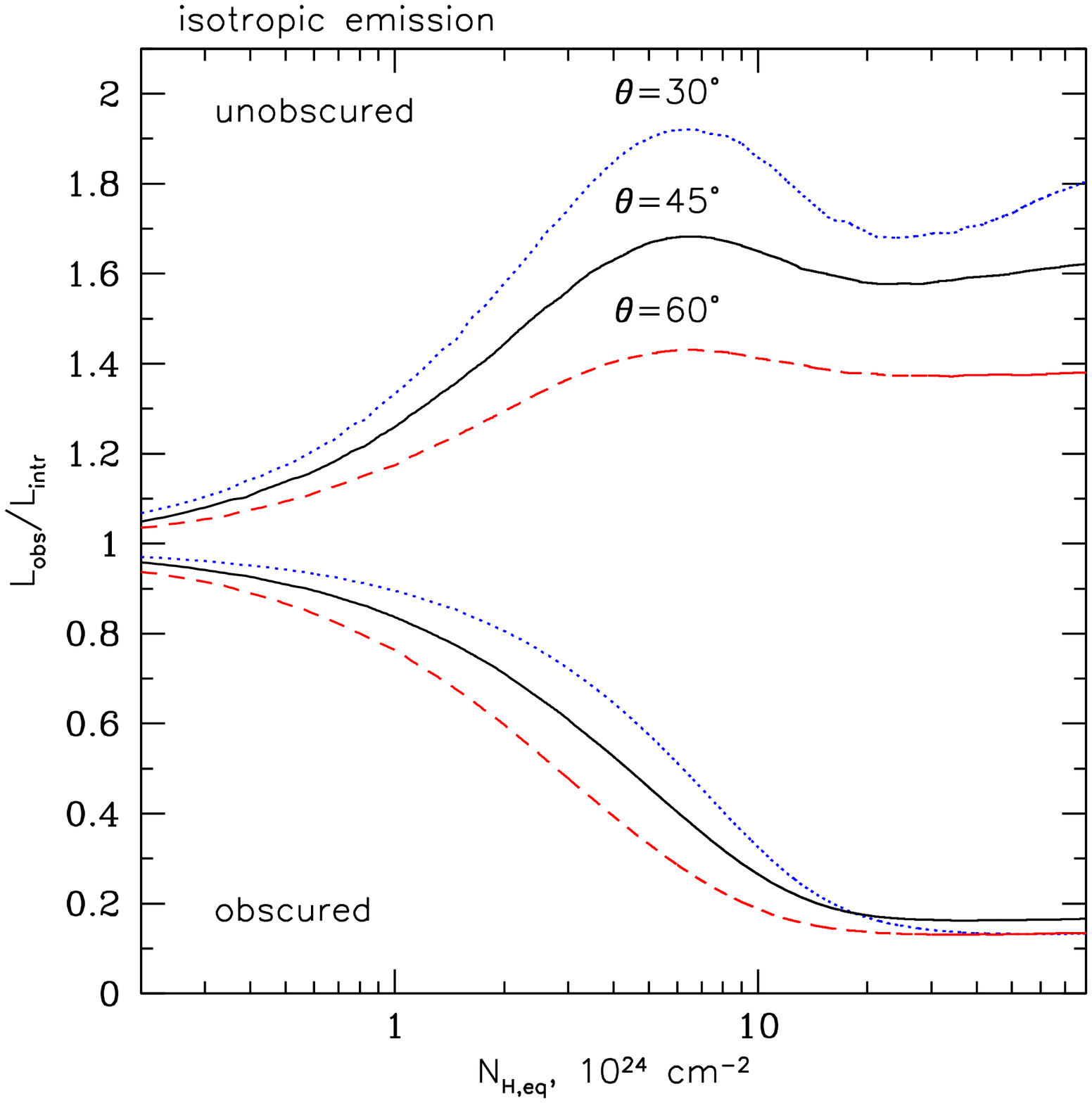}
\includegraphics[width=\columnwidth,bb=0 150 580 710]{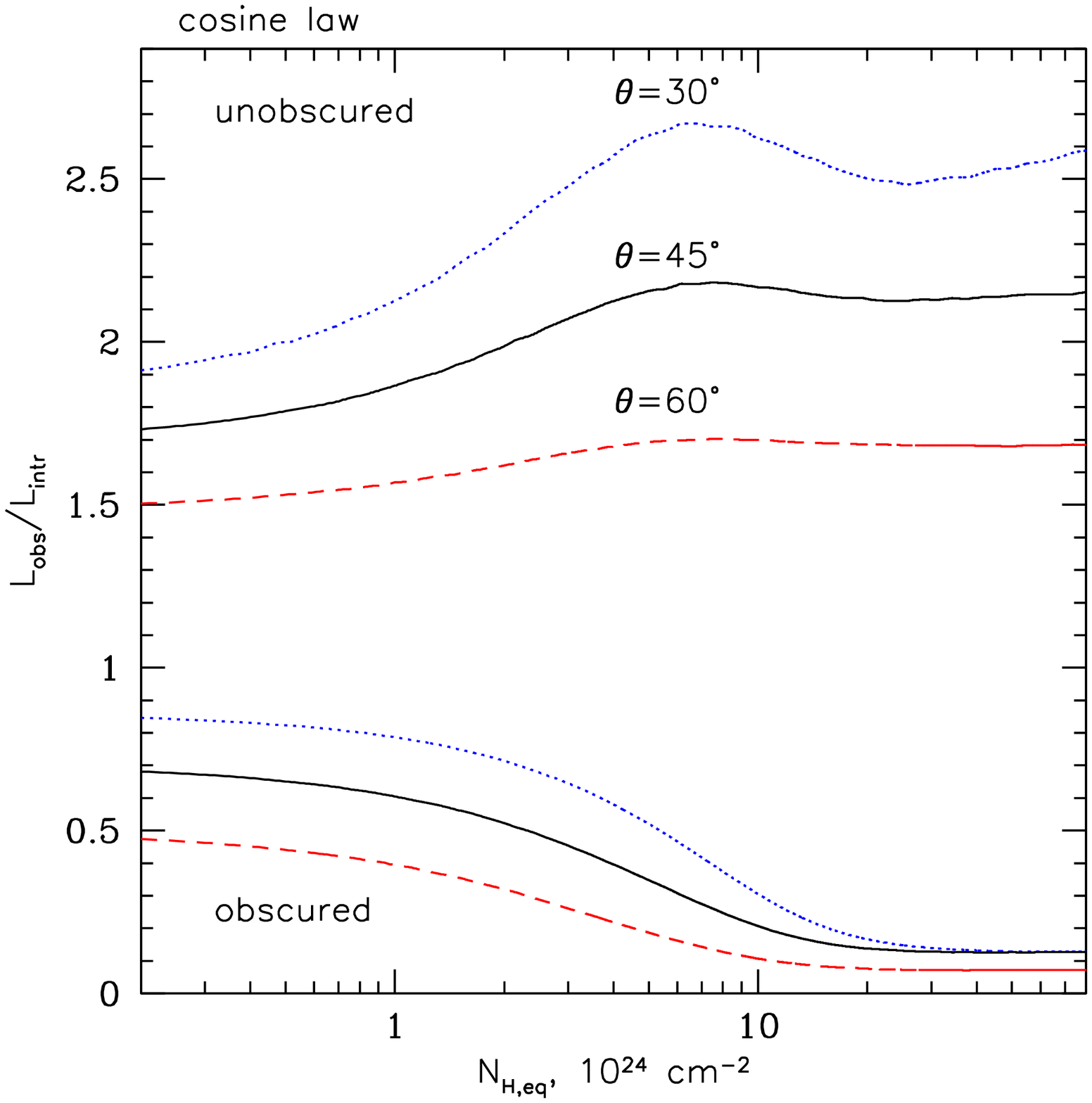}
\caption{\textit{Top:} Calculated ratio of observed to intrinsic
  (angular-averaged) luminosity in the 17--60~keV energy band averaged
  separately over all unobscured and obscured directions
  ($\alpha<\theta$ and $\alpha>\theta$, respectively) for three values
  of the torus half-opening angle, as a function of the torus column
  density, for Model~A. \textit{Bottom:} The same, but for Model~B.
}
\label{fig:flux_nh_theta}
\end{figure}

We can proceed further and ask the question: what would be the
\textit{average} observed/intrinsic flux ratio for the local 
populations of unobscured and obscured AGN if (i) AGN tori were
randomly oriented with respect to the observer, which is a 
natural assumption, and (ii) all the tori had the same half-opening
angle $\theta$ (this, of course, permits the physical size of the
torus to vary from one object to another and e.g. to depend on
luminosity). To this end, we just need to average the dependencies
shown in Fig.~\ref{fig:flux_nh_angles} over the viewing angle $\alpha$
for the unobscured and obscured directions:
\beq 
\runobsc(\nheq,\theta)=\frac{\int_{\cos\theta}^1
  R(\nheq,\theta,\alpha)d\cos\alpha}{1-\cos\theta},
\label{eq:runobsc}
\eeq
\beq
\robsc(\nheq,\theta)=\frac{\int_0^{\cos\theta}
  R(\nheq,\theta,\alpha)d\cos\alpha}{\cos\theta}.
\label{eq:robsc}
\eeq
The result is shown in Fig.~\ref{fig:flux_nh_theta} as a function of
$\nheq$ for torus half-opening angles $\theta=30^\circ$, $45^\circ$
and $60^\circ$. One can see that for Model~A, $\runobsc$ (the average
observed/intrinsic flux ratio for unobscured AGN) reaches a maximum of
$\sim 1.5$--2, depending on $\theta$, at $\nheq\sim 5\times
10^{24}$~cm$^{-2}$, then declines to $\sim 1.4$--1.7 by $\nheq\sim
1.5\times 10^{25}$~cm$^{-2}$ and stays at approximately this level for
higher column densities. The average ratio $\robsc$ for obscured AGN
monotonically decreases from $1$ to $\sim 0.2$ as $\nheq$ increases
from $\ll 10^{24}$~cm$^{-2}$ to $\sim 1.5\times 10^{25}$~cm$^{-2}$ and
stays at this level thereafter. In the case of a cosine-law emitting
source, the situation is qualitatively similar, but the contrast
between the unobscured and obscured directions is more
pronounced: it is present already at $\nh=0$ and increases further,
due to Compton reflection, with increasing $\nh$. 

It is obvious from Fig.~\ref{fig:flux_nh_theta} that a hard X-ray
survey, like the ones performed by \textit{INTEGRAL} and
\textit{Swift}, will find unobscured AGN more easily than even lightly
obscured objects, let along heavily obscured ones. Our goal now is to
correct the observed statistical properties of local AGN for this
obvious bias. 

\subsection{Intrinsic distribution of torus column densities}
\label{s:intrinsic_nh}

We can first estimate the intrinsic distribution of the column
densities, $\nheq$, of AGN torii. To this end, we need to
correct the observed $\nh$ distribution (Fig.~\ref{fig:obs_nh_distr})
for absorption bias, excluding from the consideration the first, $\nh<  
10^{22}$~cm$^{-2}$, bin since it pertains to unobscured AGN for which
our line of sight does not cross the torus. Given the fairly small
number of obscured AGN, especially of heavily obscured ones, in our
sample, we are bound to make some simplifying assumptions. For
example, we may assume that the intrinsic $\nheq$ distribution does
not depend on luminosity. In this case, the intrinsic $\nheq$
distribution can be estimated simply by dividing the observed one by
$\robsc^{3/2}(\nheq,\theta)$ (and normalising the resulting
dependence so that its integral over $\nh$ equals unity), with the
bias factor $\robsc=\lobs/\lintr$ having been discussed in
\S\ref{s:bias}.  

In doing this exercise, we assumed that $\nheq=(4/\pi)\nh\approx
1.27\nh$ for our obscured AGN. This is because we do not know the
orientation of our objects apart from the fact that some of
them are unobscured and hence $\alpha<\theta$, while others are
obscured and hence $\alpha>\theta$. For our assumed torus geometry
(see Fig.~\ref{fig:torus}), the line-of-sight column density depends
on the viewing angle as follows:
\beq
\nh(\alpha)=\nheq\sqrt{1-\left(\frac{\cos\alpha}{\cos\theta}\right)^2},
\label{eq:nh_alpha}
\eeq
so that the mean $\nh$ over all obscured directions is 
\beq
N_{\rm H,obsc}=\frac{\int_0^{\cos\theta}
  \nh(\alpha)d\cos\alpha}{\cos\theta}=\frac{\pi}{4}\nheq.
\label{eq:nh_nheq}
\eeq
Hence the coefficient in the conversion of $\nh$ to $\nheq$
above. Note that the $\nh$ values adopted from the literature for some 
of our Compton-thick AGN may already have been ascribed the meaning
of an equatorial rather than line-of-sight column density by the
corresponding authors. However, considering our sample of heavily
obscured sources as a whole, the information it contains on the
absorption columns is very heterogeneous, as it is based on various 
spectral models used by various authors. Fortunately, a typical
expected difference between $\nheq$ and $\nh$ for obscured AGN is only
$\sim 20$\% (see equation~(\ref{eq:nh_nheq}) above) and has negligible
impact on our results.  

\begin{figure}
\centering
\includegraphics[width=\columnwidth,bb=0 180 580 710]{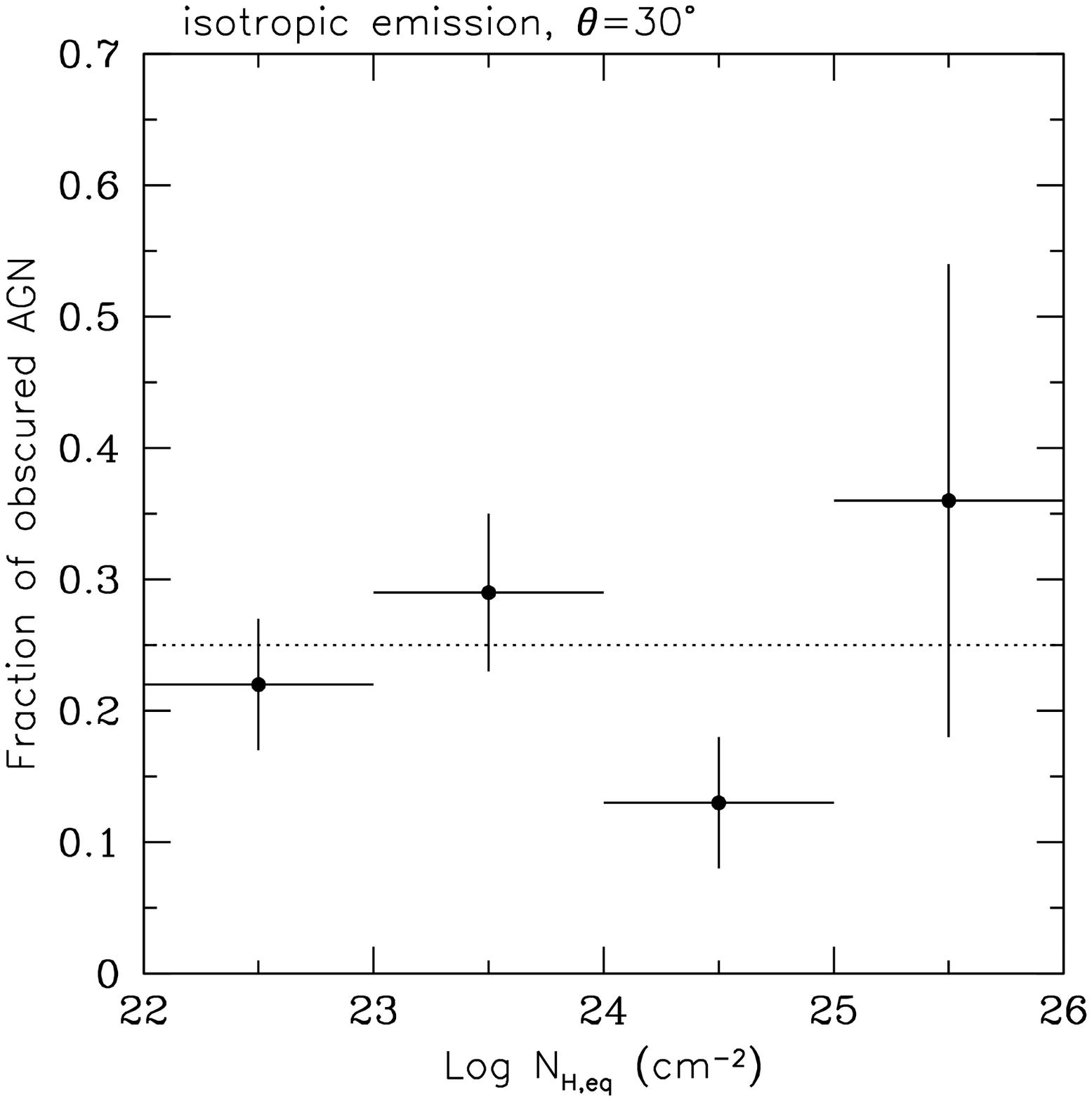}
\includegraphics[width=\columnwidth,bb=0 180 580 710]{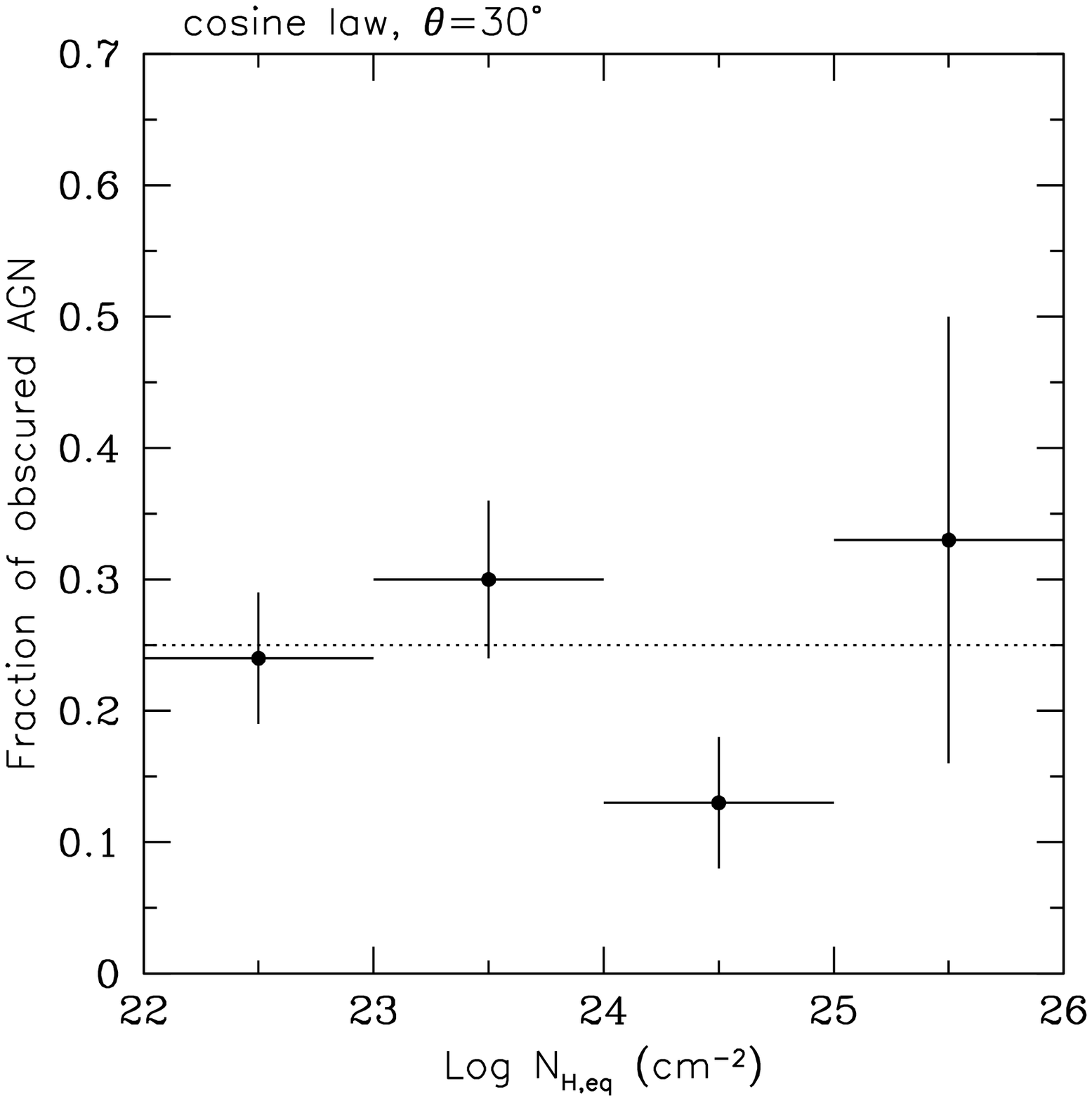}
\caption{Reconstructed intrinsic distribution of column densities of
  obscuring tori in local AGN, calculated assuming $\theta=30^\circ$
  and either Model~A (\textit{top}) or Model~B (\textit{bottom}). The
  dotted line corresponds to a log-uniform distribution.
}
\label{fig:intr_nh_distr}
\end{figure}

The resulting intrinsic $\nheq$ distribution is presented in
Fig.~\ref{fig:intr_nh_distr}. It is only weakly dependent on both the 
assumed half-opening angle $\theta$ of the obscuring torus and the
assumed emission model (Model~A or Model~B). This distribution can be 
roughly described as log-uniform between $\nheq=10^{22}$ and
$10^{26}$~cm$^{-2}$, although the upper boundary is, of course, fairly
uncertain. A similar result was previously obtained using AGN from the
\textit{Swift}/BAT hard X-ray survey
\citep{buretal11,uedetal14}. Moreover, the intrinsic $\nh$
distribution shown in Fig.~\ref{fig:intr_nh_distr} is similar to the
one inferred for optically selected Seyfert~2 galaxies
\citep{risetal99}.

\subsection{Intrinsic luminosity function}
\label{s:intrinsic_lf}

We now calculate the intrinsic hard X-ray LF of unobscured
and obscured AGN, $\phi(\lintr)\equiv dN/d\log\lintr$. As for the
observed LFs discussed in \S\ref{s:obs_properties}, we use both 
binned and analytic representations.

For the binned LFs, the procedure is as follows: 

\begin{enumerate}

\item
First, based on the observed luminosity $\lobsi$ and estimated torus column
density $\nheqi$ of each source in the sample (Table~\ref{tab:agn}), we
determine its intrinsic hard X-ray luminosity as either
$\lintri=\lobsi/\runobsc(\nheqi,\theta)$ (for unobscured sources) or
$\lintri=\lobsi/\robsc(\nheqi,\theta)$ (for obscured sources), where the
ratios $\runobsc$ and $\robsc$ are calculated as discussed above (from
eqs.~(\ref{eq:runobsc}) and (\ref{eq:robsc}), see
Fig.~\ref{fig:flux_nh_theta}), assuming some (the same for all
objects) torus half-opening angle $\theta$ and using Model~A or
Model~B. Here again we use the average ratios $\runobsc$ and $\robsc$
rather than the viewing-angle dependent $R(\nheq,\theta,\alpha)$ from
which they derive for the lack of knowlegde of the orientation of our
objects. Strictly speaking, this procedure is not fully correct,
because for given $\nheq$ and $\theta$, a hard X-ray flux limited
survey will preferentially find objects with smaller viewing angles
$\alpha$ within the corresponding groups of $\alpha<\theta$ and
$\alpha>\theta$, as is clear from
Fig.~\ref{fig:flux_nh_angles}. However, this may be regarded as a
next-order correction to the bias considered here and does not
significantly affect our results, as we verify in \S\ref{s:direct}.

As was said before, the $\nheq$ values for our obscured
objects are estimated from their measured $\nh$ columns as
$\nheq=1.27\nh$. However, we cannot determine similarly the
torus column densities for our unobscured AGN\footnote{In principle,
  one could try to estimate $\nheq$ for unobscured AGN from the
  contribution of the reflection component to the observed spectrum,
  but that requires high-quality hard X-ray data, which is not always
  available, and is model-dependent. In particular, the result will
  depend on the unknown opening angle $\theta$.}. Therefore, we simply
assume that $\nheq=10^{24}$~cm$^{-2}$ for these objects, since
this is approximately the median value of the inferred intrinsic
absorption column distribution for obscured AGN (see
Fig.~\ref{fig:intr_nh_distr}).

\item
Second, we calculate for each source the volume of the Universe,
$\vmaxi(\lobsi)$, over which AGN with such observed luminosity can be
detected in the \textit{INTEGRAL} survey. Since the detection
limit for a given hard X-ray instrument (IBIS in our case) is
actually determined by photon counts, it should depend on the
observed X-ray spectral shape, which for the problem at hand is 
affected by absorption and reflection in the torus (see examples of
AGN spectra in Fig.~\ref{fig:spectra}). We correct $\vmaxi$ for this
effect, but this correction proves to be negligible (as is the
$k$-correction due to cosmological redshift). As a result, we obtain
essentially the same $\vmaxi$ for our sources as we used in
constructing the observed LF in \S\ref{s:obs_properties}).

\item
The final step consists of summing up the $1/\vmaxi$ contributions of
the individual sources, i.e. adding the $1/\vmaxi$ for each AGN 
of a given class (unobscured or obscured) to the space density of such
objects within a luminosity bin containing $\lintri$ (rather than
$\lobsi$) for this source.

\end{enumerate}

To obtain analytic forms of the intrinsic LFs, we use the same broken
power-law model as for our observed LFs but a different likelihood
estimator:  
\beq
\mathcal{L}=-2\sum_{i}\ln\frac{\phi(\lintri)\int\vmax(\lintri,\nheq)d\log\nheq}{\int\int\phi(\lintr)\vmax(\lintr,\nheq) d\log\lintr d\log\nheq}. 
\label{eq:like_lum_intr}
\eeq
Here $\lintri$ are the same estimates of the intrinsic luminosities of
our objects as we used before to construct the binned intrinsic LFs 
(i.e. calculated from $\lobsi$ using the actual $\nhi$ estimates for the
obscured AGN and assuming that $\nheq=10^{24}$~cm$^{-2}$
for the unobscured ones), but $\vmax(\lintr,\nheq)$ is now the volume
over which AGN with given intrinsic luminosity $\lintr$ and torus
column density $\nheq$ can be detected in the \textit{INTEGRAL}
survey. To calculate these volumes, we again use the $\alpha$-averaged
quantities $\runobsc(\theta,\nheq)$ (in fitting the intrinsic LF of
unobscured AGN) and $\robsc(\theta,\nheq)$ (in fitting the intrinsic
LF of obscured AGN). The integrals over $d\log\nheq$ in
equation~(\ref{eq:like_lum_intr}) are computed from $10^{22}$ to 
$10^{26}$~cm$^{-2}$, i.e. we assume that the intrinsic distribution of
torus column densities is log-uniform over this range, as 
suggested by the result of our preceeding analysis shown in
Fig.~\ref{fig:intr_nh_distr}. 

\begin{figure}
\centering
\includegraphics[width=\columnwidth,bb=0 160 580 710]{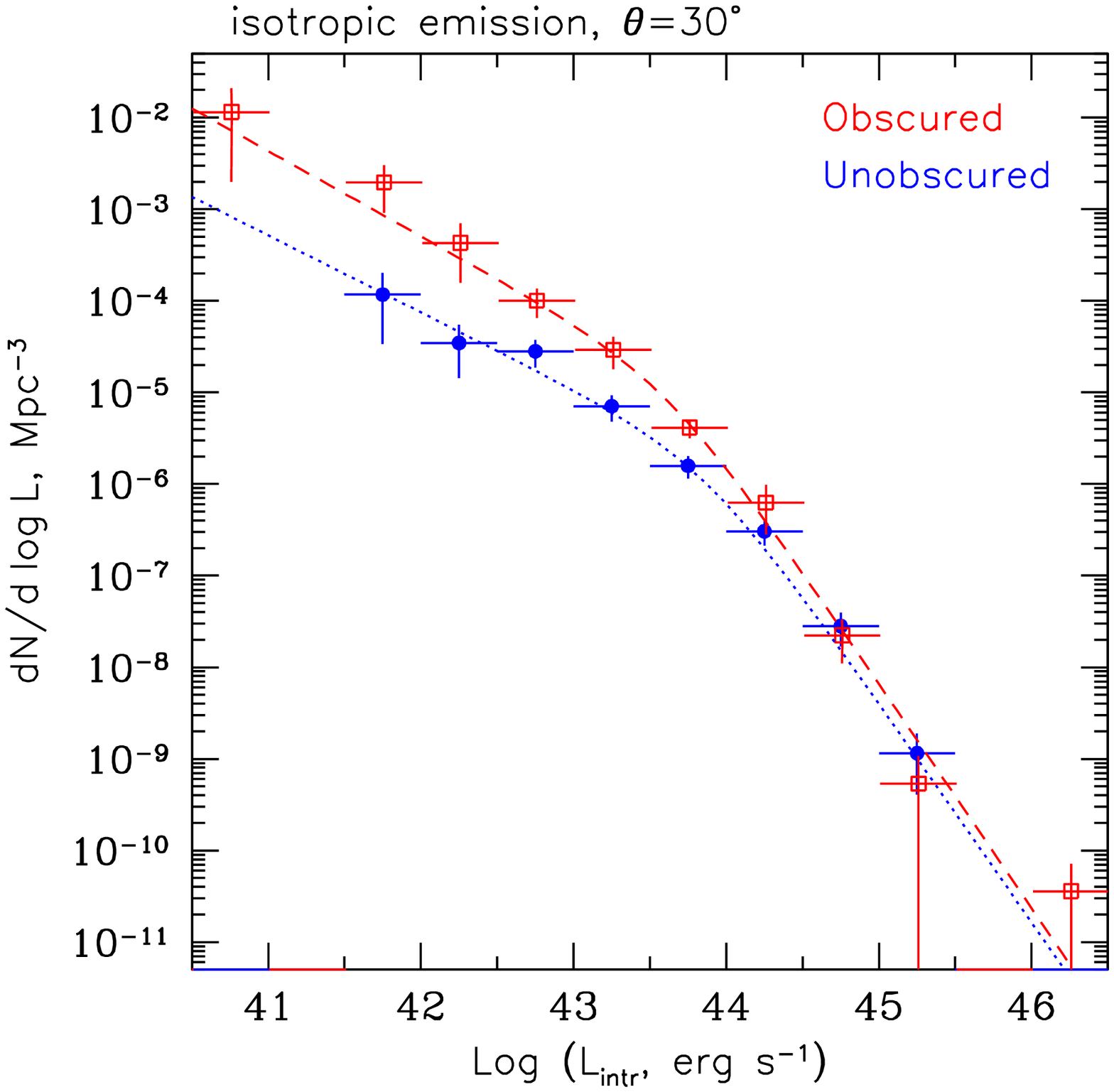}
\includegraphics[width=\columnwidth,bb=0 160 580 710]{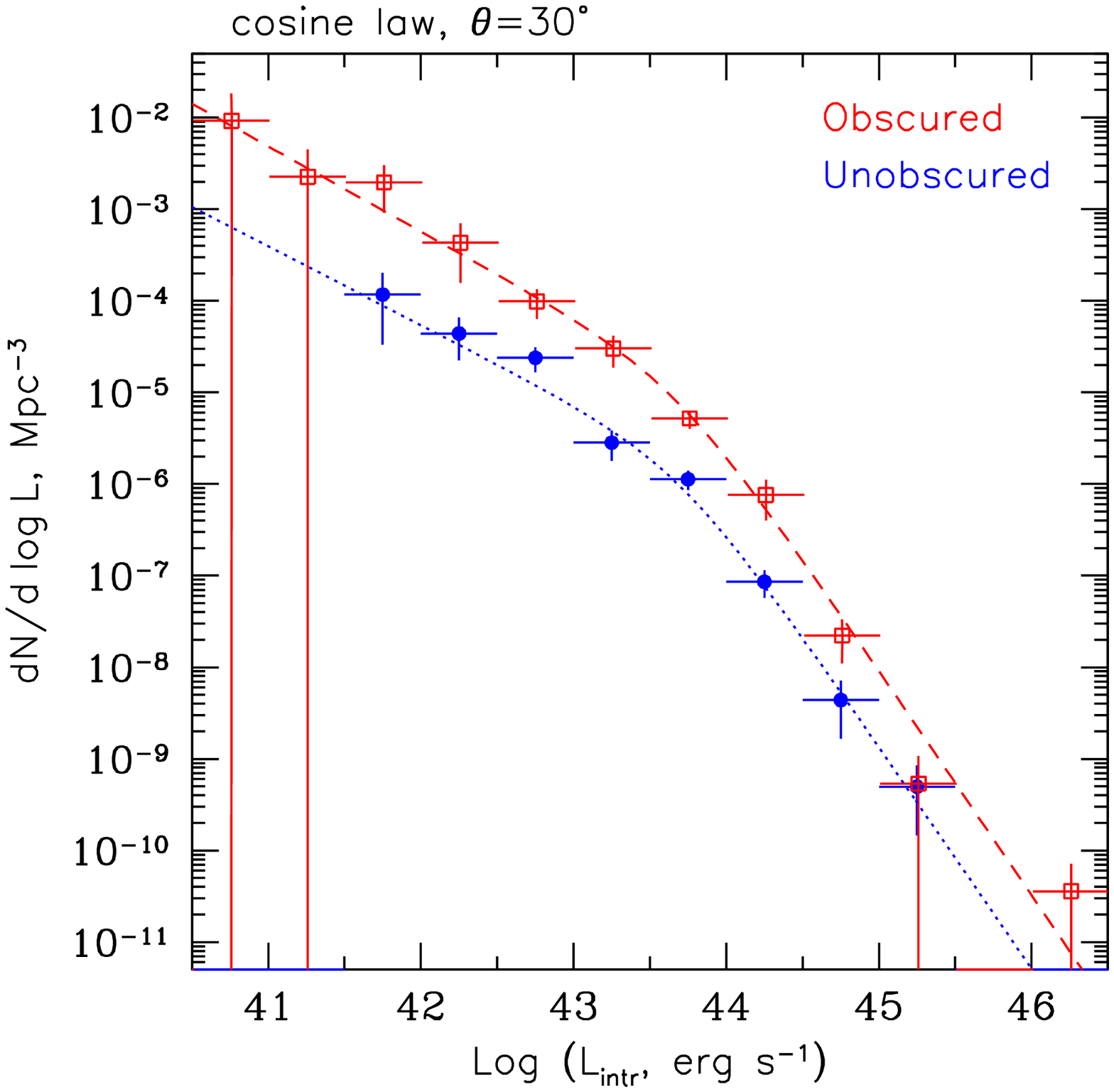} 
\caption{\textit{Top:} Reconstructed intrinsic hard X-ray LFs of
  unobscured (blue filled circles) and obscured (red empty squares)
  AGN, fitted by a broken power law (the best-fits parameters are
  given in Table~\ref{tab:lumfunc}) (blue dotted line and red dashed line,
  respectively). Model~A is adopted, with
  $\theta=30^\circ$. \textit{Bottom:} The same, but for Model~B.
}
\label{fig:intr_lumfunc_nh}
\end{figure}

\begin{figure}
\centering
\includegraphics[width=\columnwidth,bb=0 160 580 710]{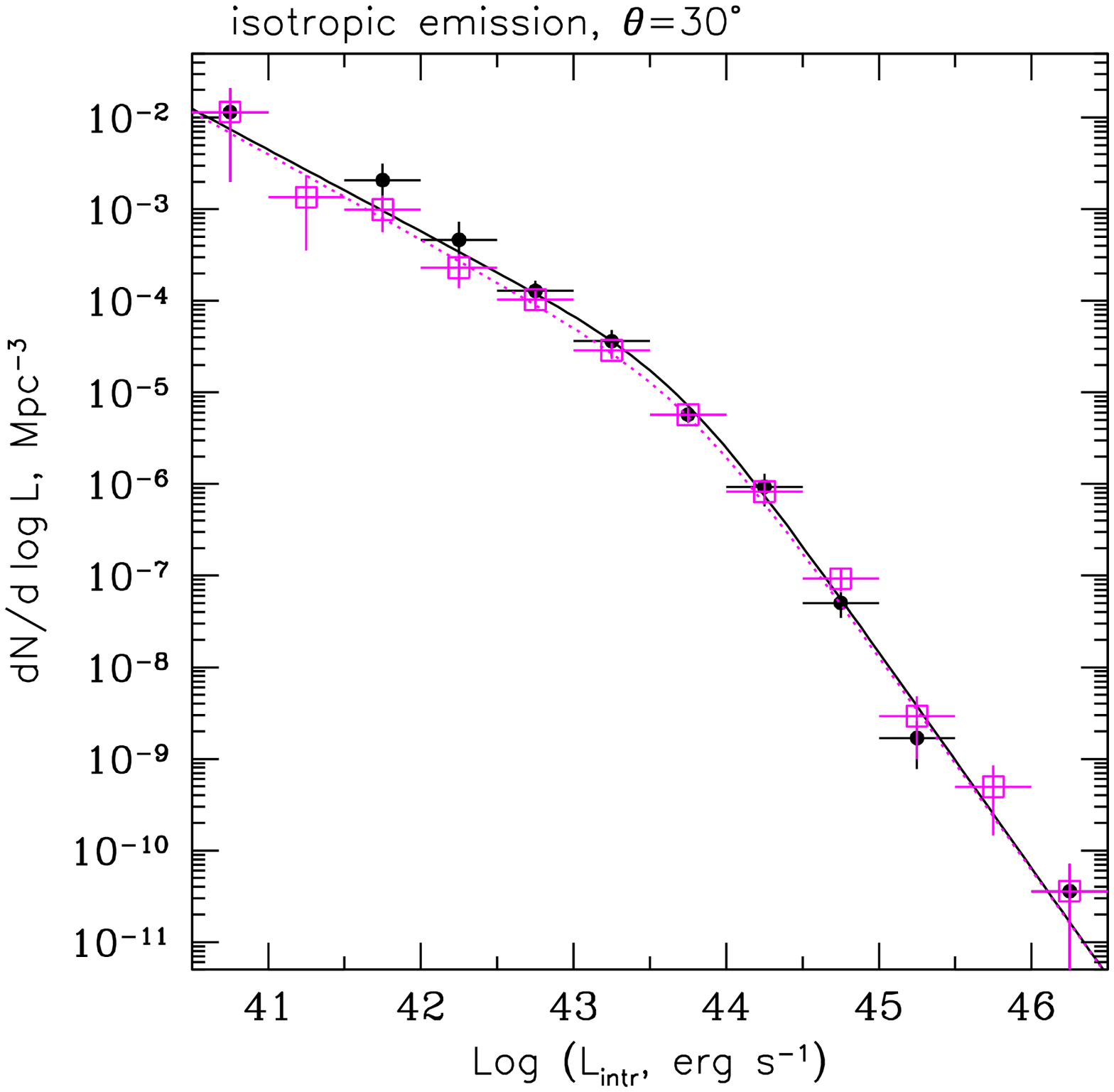}
\includegraphics[width=\columnwidth,bb=0 160 580 710]{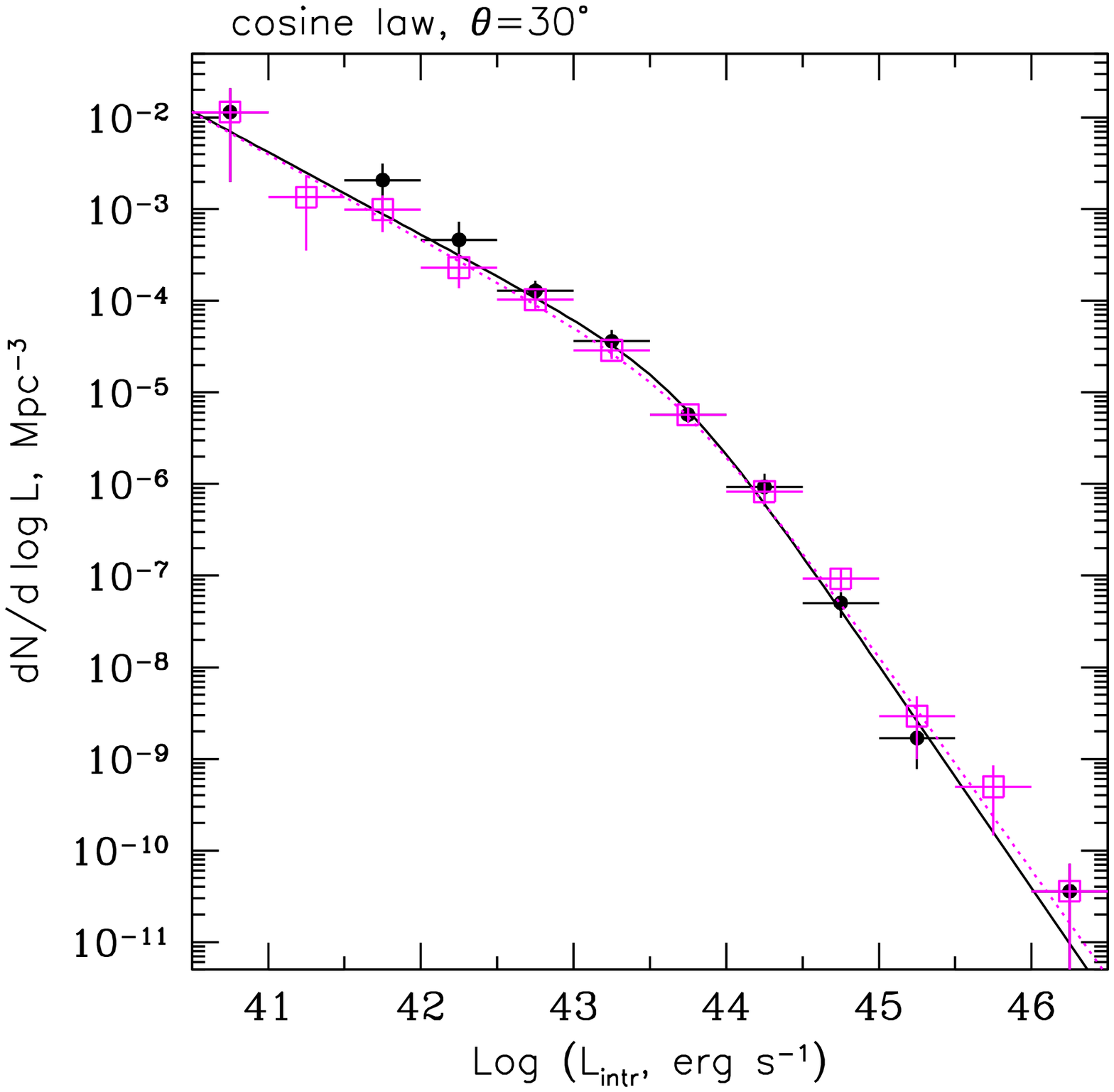}
\caption{\textit{Top:} Reconstructed intrinsic hard X-ray LF of local AGN (black
  filled circles), fitted by a broken power law (black solid line, the
  best-fit parameters are given in Table~\ref{tab:lumfunc}). Model~A
  is adopted, with $\theta=30^\circ$. For comparison, the observed LF
  is also shown (magenta empty squares and dotted
  line). \textit{Bottom:} The same, but for Model~B.
}
\label{fig:intr_lumfunc}
\end{figure}

Figure~\ref{fig:intr_lumfunc_nh} shows the resulting intrinsic LFs for
unobscured and obscured AGN, calculated assuming $\theta=30^\circ$ for
Model~A and Model~B. We see that in the former case, the shapes of the
intrinsic LFs of unobscured and obscured AGN are clearly different
from each other, although to a lesser degree that it was for the
observed LFs (Fig.~\ref{fig:obs_lumfunc_nh}) from which they 
derive. However, for Model~B the intrinsic LFs of unobscured and
obscured AGN are not significantly different in shape from each
other. These conclusions are verified by the best-fit parameters
obtained for these LFs (see Table~\ref{tab:lumfunc}). Note that the derived
intrinsic LFs (both binned and analytic ones) are only weakly
sensitive to the torus half-opening angle $\theta$ that was assumed in
constructing them, and nearly the same results are obtained for
$\theta=30^\circ$ and $\theta=45^\circ$. This is due to the weak
sensitivity of the $\runobsc$ and $\robsc$ factors to $\theta$ (see
Fig.~\ref{fig:flux_nh_theta}).

The transformation of the observed binned LFs to the intrinsic ones
can be understood as follows: (i) all unobscured AGN making up the LF
shift by the same amount, $\log\lobs/\lintr\sim 0.1$ and $\sim 0.3$ for
Model~A and Model~B, respectively, to the left along the luminosity
axis (since we have assumed the same equatorial optical depth of the
torus, $\nheq=10^{24}$~cm$^{-2}$, for all of our unobscured objects),
and (ii) each obscured AGN from the \textit{INTEGRAL} sample moves its
own distance to the right-hand side of the plot, this shift being
small for lightly obscured objects ($\nh<10^{24}$~cm$^{-2}$) but
substantial (up to $\log\lintr/\lobs\sim 1$ for $\nh\gtrsim
10^{25}$~cm$^{-2}$) for heavily obscured ones.

Finally, we can calculate the intrinsic LF of the entire local AGN
population, by summing up the contributions of unobscured and obscured
sources. In obtaining the analytic fit in this case, we define  
$\vmax(\lintr,\nheq)$ as follows:
\beq
\vmax=\vmaxunobsc(1-\cos\theta)+\vmaxobsc\cos\theta,
\eeq
where $\vmaxunobsc(\lintr,\nheq)$ and $\vmaxobsc(\lintr,\nheq)$ are
the corresponding volumes for unobscured and obscured AGN. The
resulting LF is shown in Fig.~\ref{fig:intr_lumfunc} and 
its best-fit parameters are presented in Table~\ref{tab:lumfunc}. 

One can see that the total intrinsic LF is not very different from
the total observed LF. This means that the two effects observed in
Fig.~\ref{fig:intr_lumfunc_nh}, namely the shift of the LF of
unobscured AGN to lower luminosities and the shift of the LF of
obscured AGN to higher luminosities almost compensate each
other, with this conclusion being only weakly sensitive to the assumed
torus opening angle and angular dependence of intrinsic emission.

\subsection{Total AGN space density}
\label{s:intrinsic_dens}

Integration of the total intrinsic and observed LFs over luminosity
suggests that the cumulative hard X-ray luminosity density of local
AGN may be underestimated by the observed LF by $\sim 10$--30\%,
although this increase is statistically insignificant (see
Table~\ref{tab:lumfunc}). Specifically, the intrinsic luminosity
density of AGN with $\lintr>10^{40.5}$~erg~s$^{-1}$ is found to be 
$\sim 1.8\times 10^{39}$~erg~s$^{-1}$~Mpc$^{-3}$ (17--60~keV), with
the exact value slightly depending on our assumptions (see
Table~\ref{tab:lumfunc}). 
 
For our assumed intrinsic AGN spectrum ($dN/dE\propto
E^{-1.8}e^{-E/200~{\rm keV}}$), the ratio of luminosities in the
2--10~keV and 17--60~keV energy bands is about unity. Therefore, the
luminosity density of AGN with $\lintr>10^{40.5}$~erg~s$^{-1}$ may be
estimated at $\sim 1.8\times 10^{39}$~erg~s$^{-1}$~Mpc$^{-3}$ also in
the standard X-ray band (2--10~keV). We may compare this value with a 
prediction for $z=0$ based on a redshift-dependent intrinsic
luminosity function derived by \cite{uedetal14} using a large
heterogenous sample of AGN compiled from various surveys. Integration
of this LF over the luminosity range from $10^{40.5}$ to
$10^{46.5}$~erg~s$^{-1}$ gives $\sim 8\times
10^{38}$~erg~s$^{-1}$~Mpc$^{-3}$ (2--10~keV), which is a factor of
$\sim 2$ smaller than the above estimate. In reality, the
$\lintr(17-60~{\rm keV})$/$\lintr(2-10~{\rm keV})$ ratio may well be
$\sim 1.5$ rather than $\sim 1$ due to the expected presence in AGN
spectra of a Compton reflection component associated with the
accretion disk. In fact, this component was already discussed in
\S\ref{s:torus} as one of the reasons why hard X-ray emission may
be intrinsically collimated in AGN and is implicitly taken into account
in our anisotropic Model~B. Taking this spectral component into
account, we can lower our estimate of the luminosity density to 
$\sim 1.2\times 10^{39}$~erg~s$^{-1}$~Mpc$^{-3}$ (2--10~keV), which is
still higher than the \cite{uedetal14} result by a factor of
$\sim 1.5$. The remaining difference may be related to the
different procedures used in these works to construct the intrinsic
LFs and to the larger and more complete sample of local heavily
obscured AGN used in our study. 

\subsection{Intrinsic dependence of obscured AGN fraction on luminosity}
\label{s:intrinsic_frac}

\begin{figure}
\centering
\includegraphics[width=0.85\columnwidth,bb=0 160 580 710]{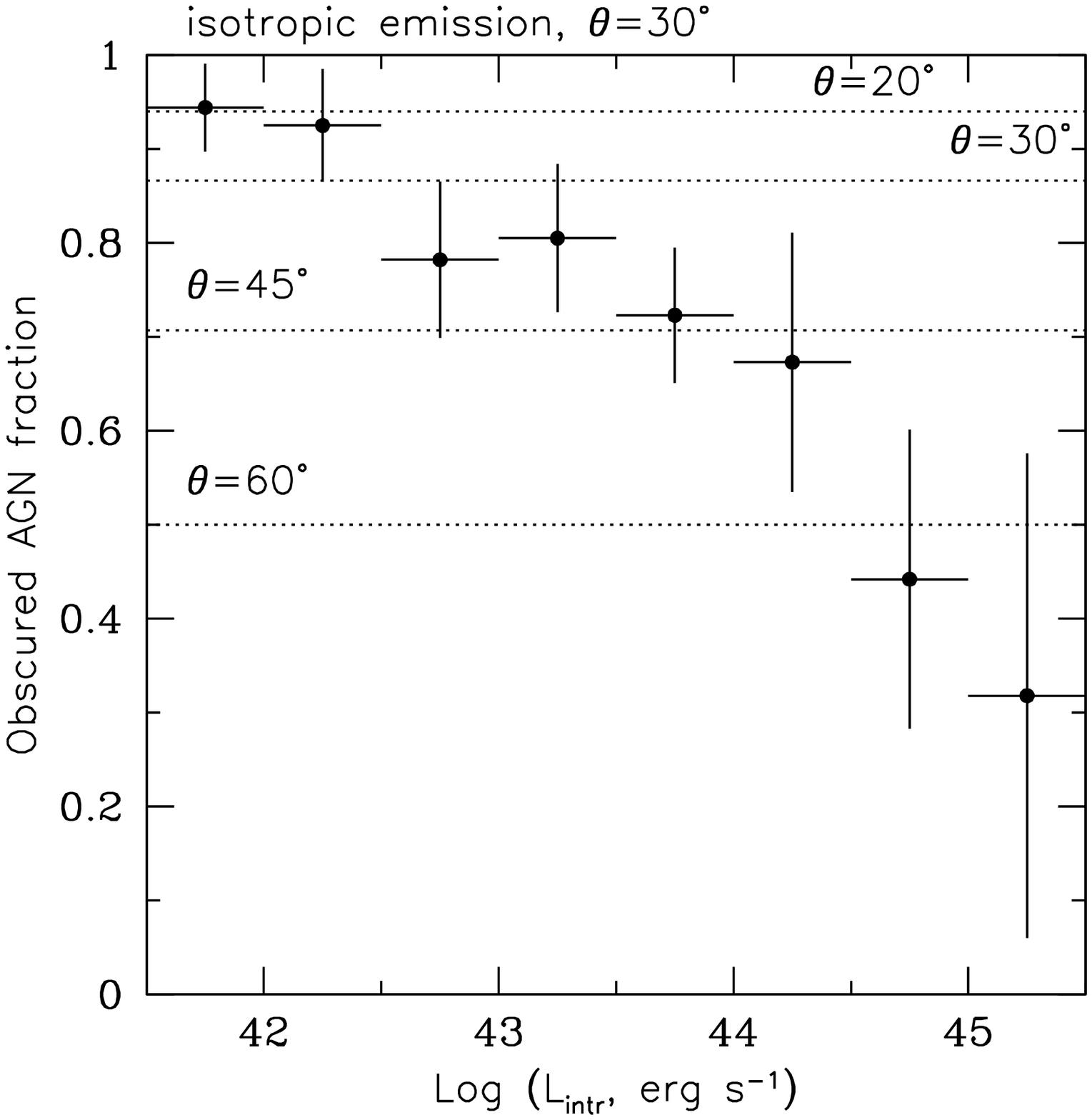}
\includegraphics[width=0.85\columnwidth,bb=0 160 580 710]{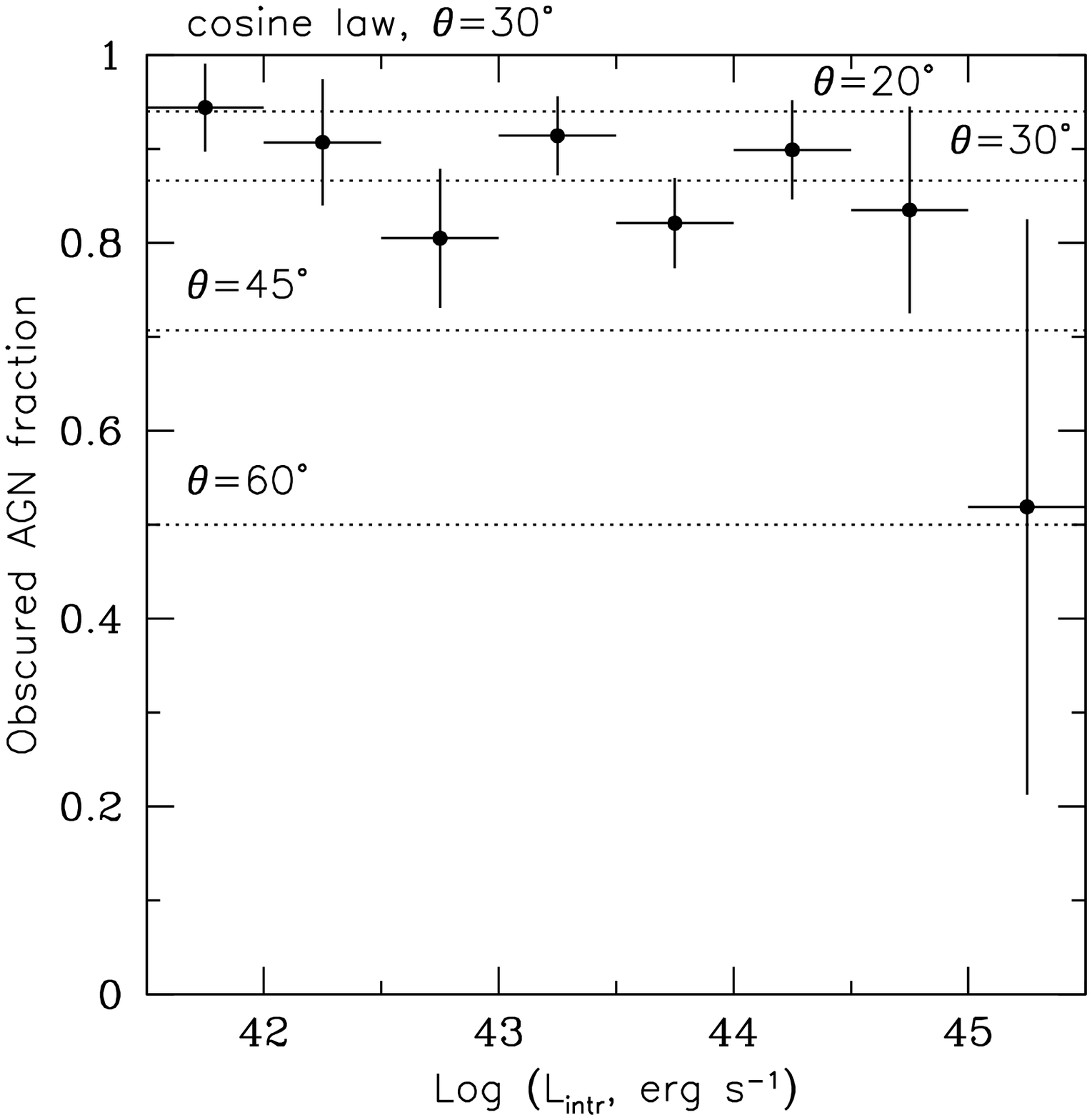}
\caption{\textit{Top:} Reconstructed intrinsic fraction of obscured
  AGN as a function of intrinsic hard X-ray luminosity, calculated for
  $\theta=30^\circ$ and Model~A. \textit{Bottom:} The same but for
  Model~B. The dotted lines indicate the fraction of the sky
  that will be screened from the central source by a torus with
  half-opening angle $\theta=20^\circ$, $30^\circ$, $45^\circ$ or
  $60^\circ$. 
}
\label{fig:intr_obsc_frac_30}
\end{figure}

\begin{figure}
\centering
\includegraphics[width=0.85\columnwidth,bb=0 160 580 710]{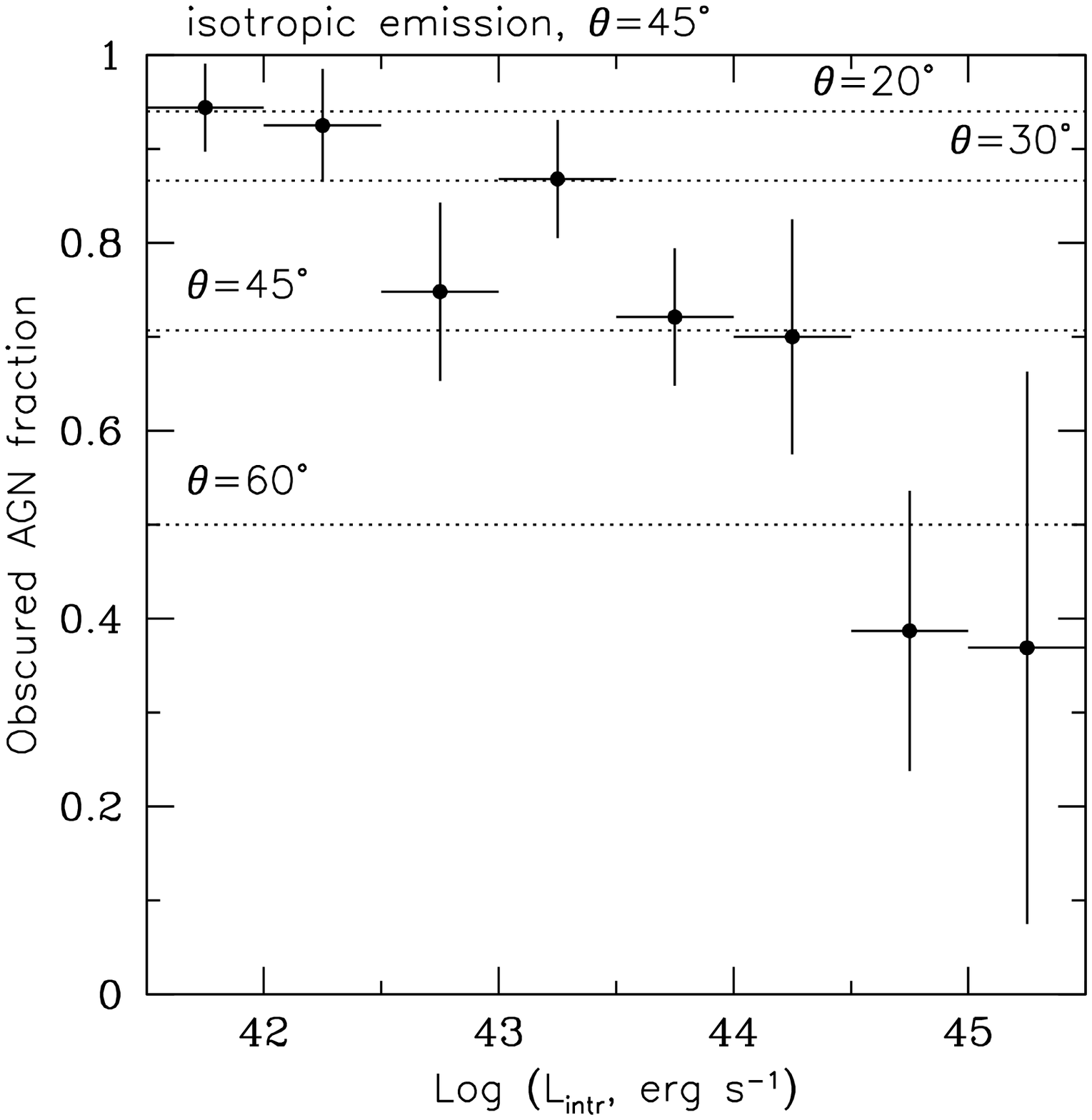}
\includegraphics[width=0.85\columnwidth,bb=0 160 580 710]{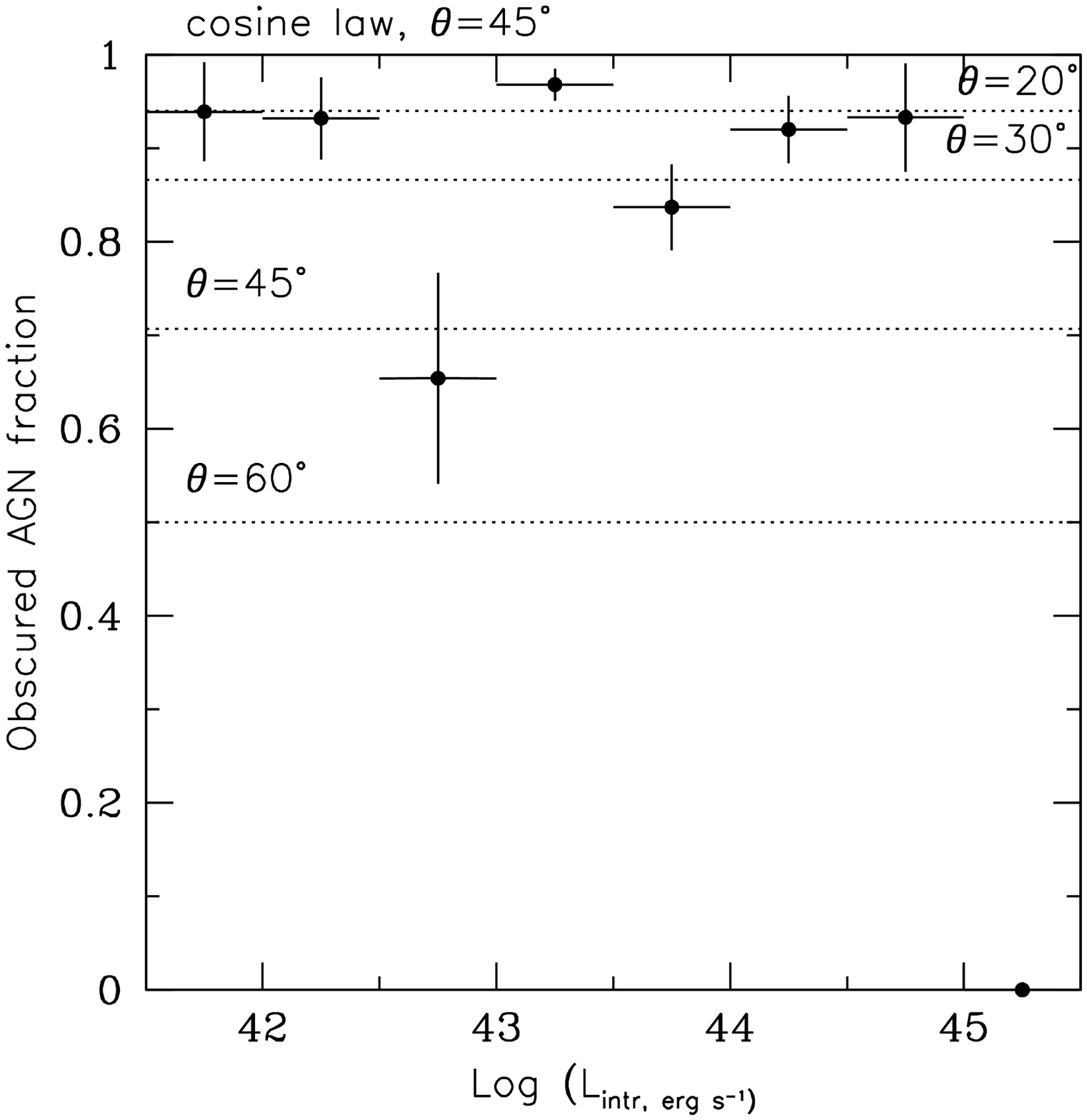}
\caption{The same as Fig.~\ref{fig:intr_obsc_frac_30}, but for
  $\theta=45^\circ$.
}
\label{fig:intr_obsc_frac_45}
\end{figure}

Similarly to the observed LF, the observed dependence of the fraction
of obscured AGN on luminosity (Fig.~\ref{fig:obs_obsc_frac}) must be
affected by AGN detection biases. By removing these biases one can 
obtain an intrinsic dependence of the obscured AGN fraction on
luminosity. To this end, we should simply divide the intrinsic LF 
of obscured AGN by the total intrinsic LF. The result is
presented in Fig.~\ref{fig:intr_obsc_frac_30} for $\theta=30^\circ$,
Model~A and Model~B, and in Fig.~\ref{fig:intr_obsc_frac_45} for
$\theta=45^\circ$.

We see that in the case of isotropic emission the declining
trend of obscured AGN fraction with luminosity is retained upon
removing the absorption bias, although the intrinsic obscured fraction
at any luminosity is significantly higher compared to the observed
fraction (see Fig.~\ref{fig:obs_obsc_frac}). We can interpret this
result in terms of the torus opening angle, i.e. the fraction of the
sky that will be shielded from the central source by a toroidal
structure of gas. To this end, we have drawn in
Figs.~\ref{fig:intr_obsc_frac_30} and \ref{fig:intr_obsc_frac_45} four
horizontal lines corresponding 
to $\theta=20^\circ$, $30^\circ$, $45^\circ$ and $60^\circ$. We see that 
if the central sources in AGN are isotropic, then the torus opening
angle must be smaller than $30^\circ$ in low-luminosity objects 
($\lintr\lesssim 10^{42.5}$~erg~s$^{-1}$) and increasing to $\sim
45^\circ$--$60^\circ$ in high-luminosity ones ($\lintr\gtrsim
10^{44}$~erg~s$^{-1}$).  

If, however, the emission from the central SMBH is collimated as
$d\lintr/d\Omega\propto\cos\alpha$, then the derived 
intrinsic dependence of the obscured AGN fraction on luminosity is in
fact consistent with the opening angle of the torus being constant with
luminosity, namely $\theta\sim 30^\circ$ -- see the bottom panel
in Fig.~\ref{fig:intr_obsc_frac_30}.

Importantly, these conclusions are almost insensitive to the opening
angle (within the range $\theta\sim 20$--$60^\circ$) of the torus that we
actually assumed in deriving the intrinsic luminosity dependences of
the obscured AGN fraction (compare the results for $\theta=45^\circ$
and $\theta=30^\circ$ in Figs.~\ref{fig:intr_obsc_frac_30} and 
\ref{fig:intr_obsc_frac_45}). This is again due to the fact that the 
$\runobsc$ and $\robsc$ ratios, which characterise observational
biases for our AGN sample and which we have corrected for, are, for a
given emission law (Model~A or Model~B), primarily determined by the
torus column density (see Fig.~\ref{fig:flux_nh_theta}). Hence, the
derived intrinsic luminosity dependences of the obscured AGN fraction
are quite robust but depend on the actual degree of collimation of
the AGN central source.

\section{Direct convolution model}
\label{s:direct}

We have demonstrated that positive bias with respect to unobscured AGN
and negative bias with respect to obscured AGN in flux-limited hard X-ray
surveys together strongly affect the observed dependence of the
obscured AGN fraction on luminosity. Our preceeding analysis consisted
of convering the observed LFs of unobscured and obscured AGN to the 
intrinsic LFs of these populations. In solving this 'inverse problem',
we used a number of simplifications that we noted were unlikely to
have significant impact on our results. In particular, we used the
viewing angle-averaged conversion factors $\runobsc(\lintr,\nheq)$ and
$\robsc(\lintr,\nheq)$ rather than the $\runobsc(\lintr,\nheq,\alpha)$
and $\robsc(\lintr,\nheq,\alpha)$ ratios from which they derive. We also
assumed a fixed torus column density, $\nheq=10^{24}$~cm$^{-2}$, 
for all of our unobscured AGN. To verify that these assumptions were
reasonable, we now perform a 'direct convolution' test, as
described below: 

\begin{enumerate}

\item
  Assume that the tori in local AGN have the same half-opening angle
  $\theta$. 

\item
  Assume that local AGN are oriented randomly with respect to us.

\item
Assume that the AGN central source is isotropic or, alternatively,
emitting according to Lambert's law ($d\lintr/d\Omega\propto\cos\alpha$). 

\item
Assume that the intrinsic distribution of AGN torus column densities
$\nheq$ does not depend on luminosity and is log-uniform 
between $\nhmin$ and $\nhmax$. Such a distribution, with $\nhmin\sim
10^{22}$~cm$^{-2}$ and $\nhmax\sim10^{26}$~cm$^{-2}$, approximately
matches the real $\nheq$ distribution we have inferred using the
\textit{INTEGRAL} sample (see Fig.~\ref{fig:intr_nh_distr}). 

\item
Adopt the intrinsic AGN LF as derived in our preceeding analysis for 
given $\theta$ and emission law (see Table~\ref{tab:lumfunc}).

\item
Use the above set of assumptions specifying the intrinsic properties
of the local AGN population to simulate, using our torus obscuration
model, AGN properties as would be observed in the \textit{INTEGRAL}
survey. 

\end{enumerate}

The main difference with respect to the inverse problem is that the
$\lobs/\lintr$ ratio now explicitly depends on the viewing angle,
which is randomly drawn for each simulated source, and on $\nheq$,
which is drawn from the assumed log-uniform distribution for each
simulated source, both for obscured AGN and for unobscured ones.

\begin{figure}
  \centering
  \includegraphics[width=0.95\columnwidth,bb=0 180 580 710]{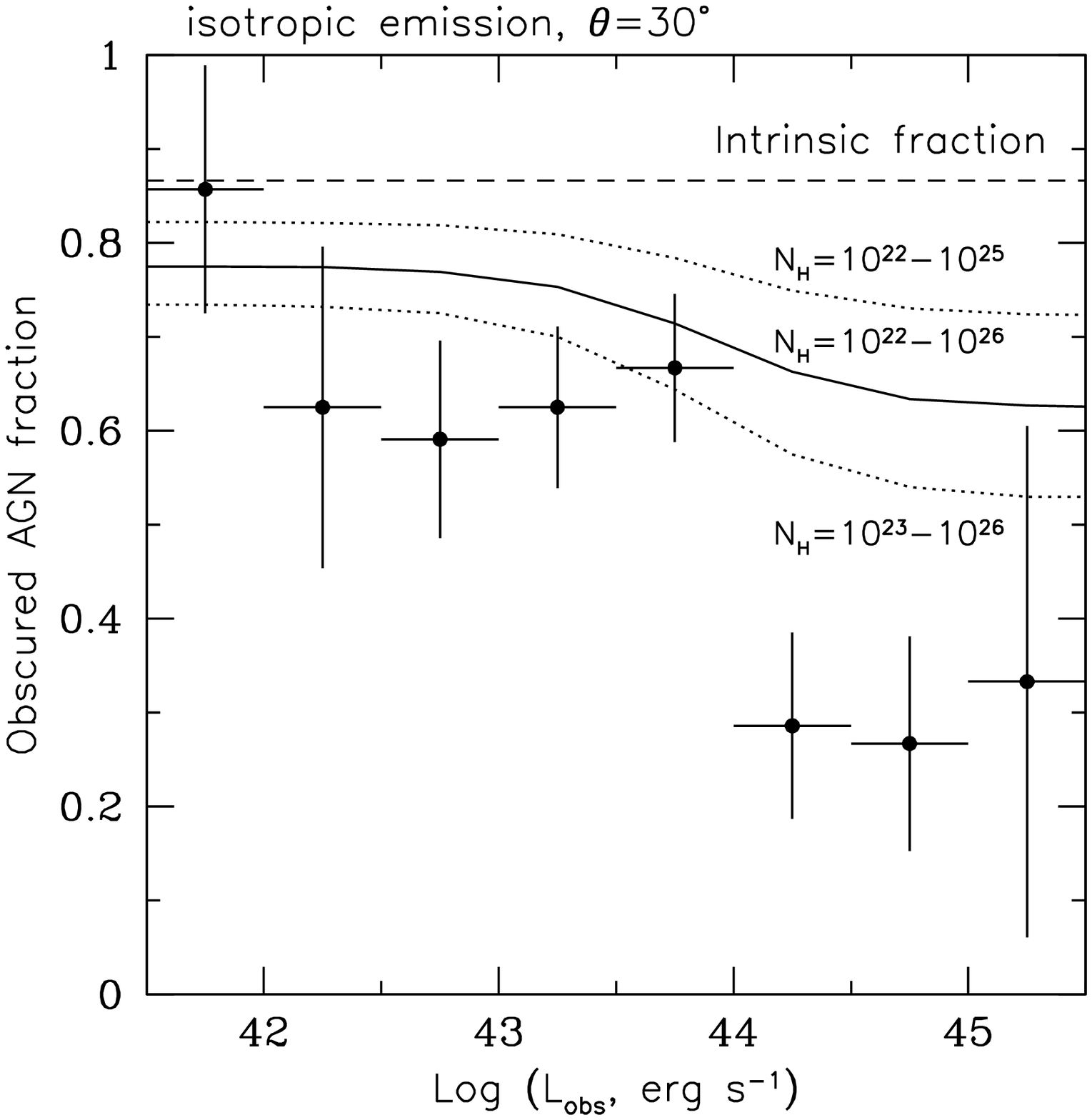}
  \includegraphics[width=0.95\columnwidth,bb=0 180 580 710]{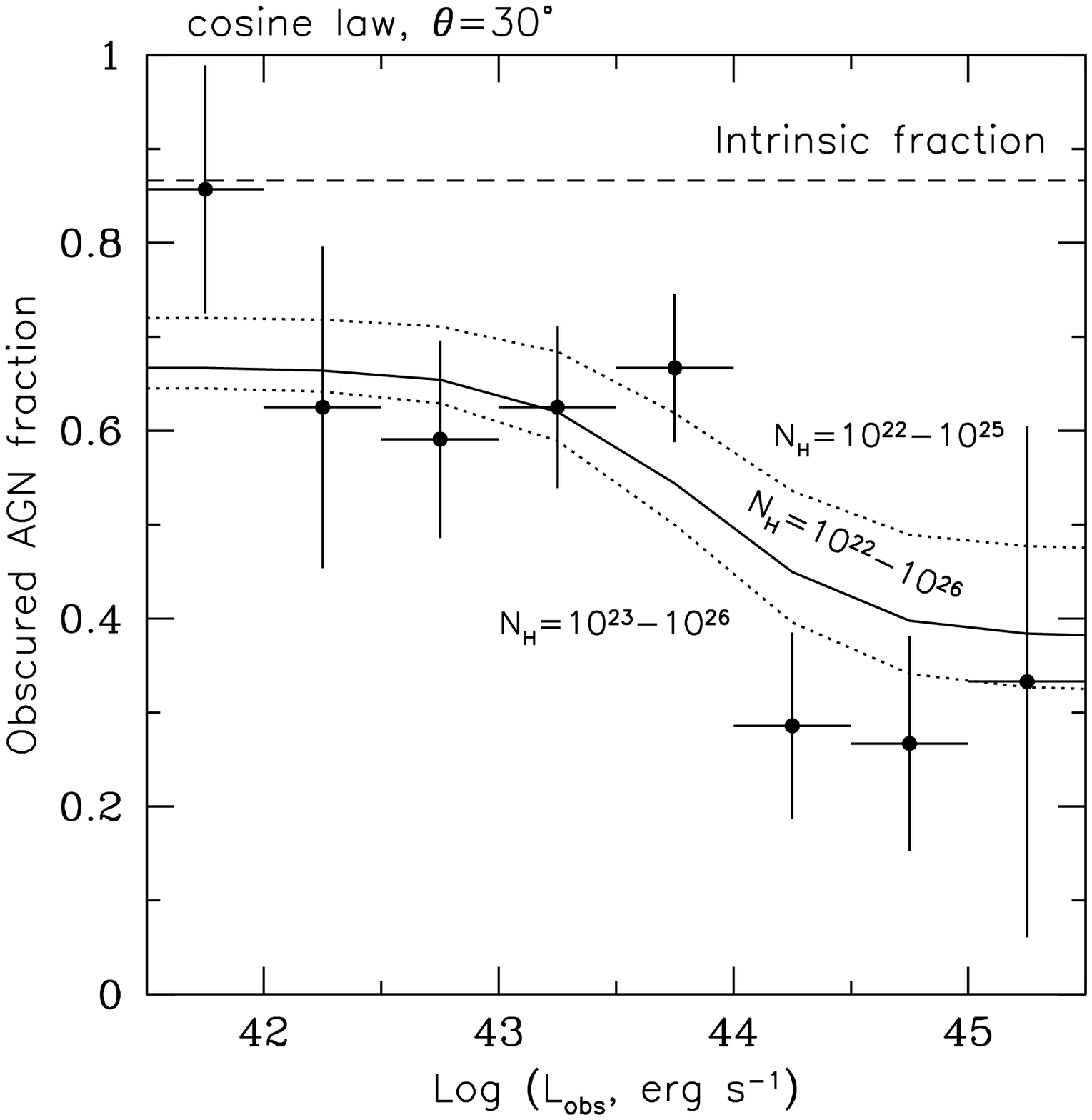}
  \caption{Simulated dependence of the observed fraction of obscured
    AGN on observed hard X-ray luminosity for $\theta=30^\circ$ and
    either isotropic (\textit{top}) or cosine-law (\textit{bottom})
    emission, for various ranges of $\nheq$ (solid and dotted
    lines). The intrinsic obscured fraction is indicated by the dashed
    line. For comparison, the corresponding dependence observed with
    \textit{INTEGRAL} is reproduced from Fig.~\ref{fig:obs_obsc_frac}
    (data points with error bars).
  }
  \label{fig:toymodel_30}
\end{figure}

\begin{figure}
  \centering
  \includegraphics[width=0.95\columnwidth,bb=0 180 580
  710]{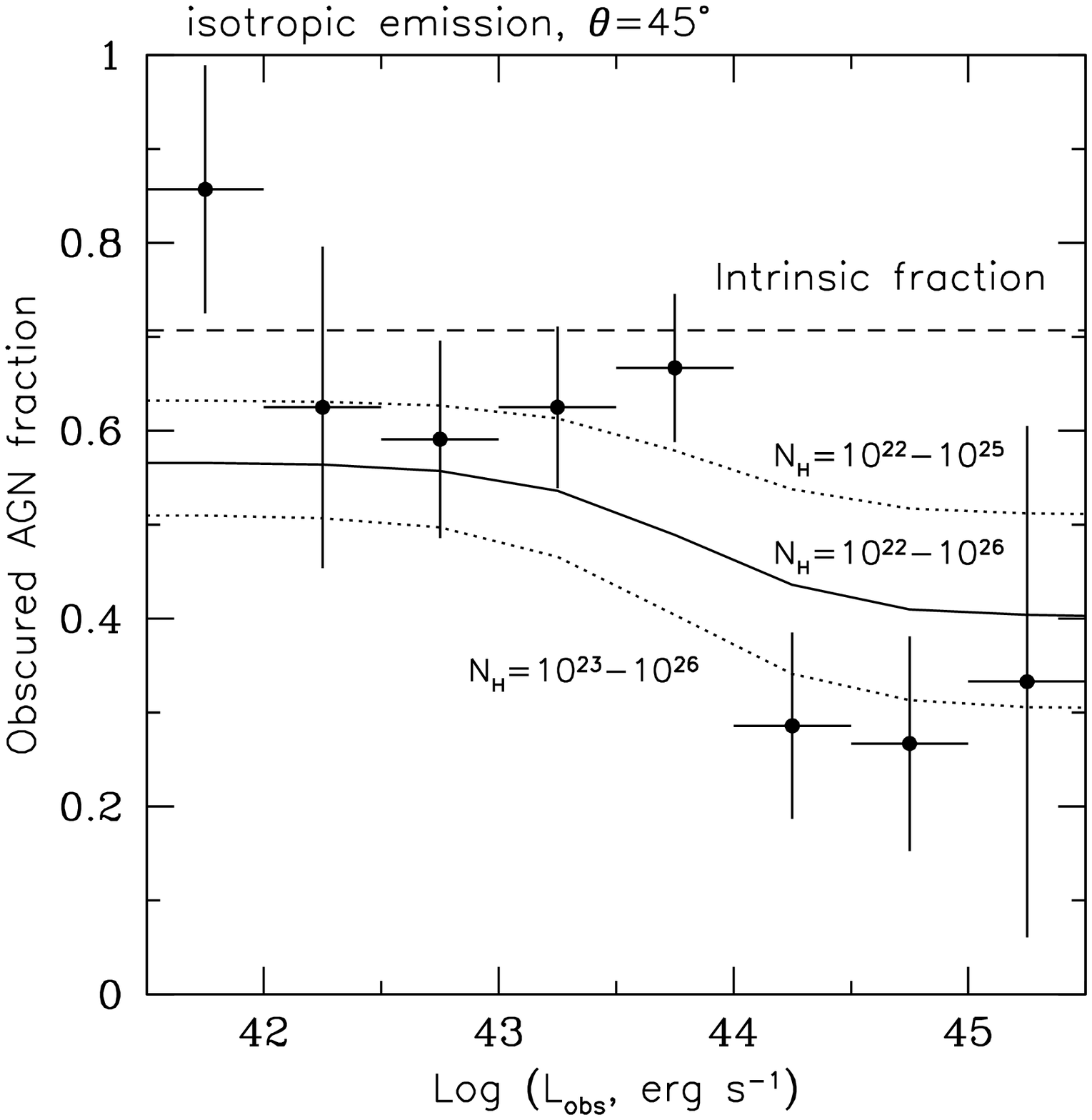}
  \includegraphics[width=0.95\columnwidth,bb=0 180 580
  710]{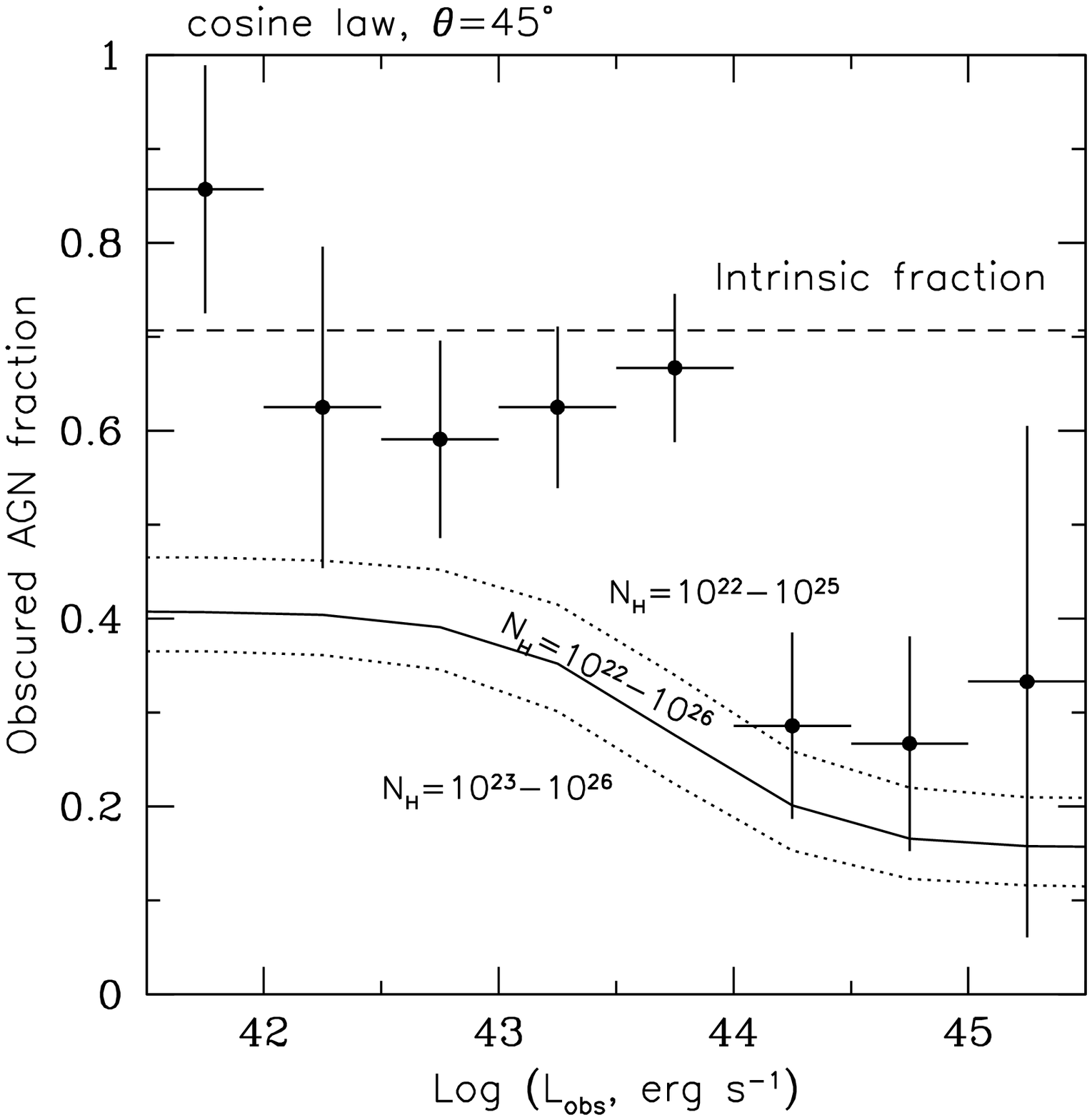}
  \caption{The same as Fig.~\ref{fig:toymodel_30}, but for
    $\theta=45^\circ$.
  }
  \label{fig:toymodel_45}
\end{figure}

Figure~\ref{fig:toymodel_30} shows the simulated luminosity
dependences of the observed obscured AGN fraction for
$\theta=30^\circ$ and either isotropic or cosine-law emission;
Fig.~\ref{fig:toymodel_45} shows the corresponding results for
$\theta=45^\circ$. As our baseline $\nheq$ distribution we use
$\nhmin=10^{22}$~cm$^{-2}$ and $\nhmax=10^{26}$~cm$^{-2}$ (solid
lines), but we also show results obtained for
$\nhmin=10^{22}$~cm$^{-2}$ and $\nhmax=10^{25}$~cm$^{-2}$ and for
$\nhmin=10^{23}$~cm$^{-2}$ and $\nhmax=10^{26}$~cm$^{-2}$. The results
of simulations are compared with the luminosity dependence actually
observed with \textit{INTEGRAL}.

We see that the luminosity dependence of the observed obscured AGN
fraction predicted for the case of isotropic emission and
$\theta=30^\circ$ is inconsistent with the \textit{INTEGRAL} data
($\chi^2=32.6$ per 8 data points between $\lobs=10^{41.5}$ and
$10^{45.5}$~erg~s$^{-1}$). The isotropic model with a larger torus
opening angle, $\theta=45^\circ$, provides a better match but the fit
is nevertheless poor ($\chi^2=15.2$ per 8 data points). Among the four
presented cases, the best agreement between simulations and
observations is achieved in the case of cosine-law emission and
$\theta=30^\circ$ ($\chi^2=9.0$ per 8 data points). These results
confirm our previously reached conclusion that unless hard X-ray
emission in AGN is intrinsically collimated, there must be an
intrinsic declining trend of the torus opening angle with increasing
AGN luminosity. 

Nevertheless, Figs.~\ref{fig:toymodel_30} and \ref{fig:toymodel_45}
clearly demonstrate that biases associated with detection of AGN in
flux-limited hard X-ray surveys \textit{inevitably lead to the
  observed fraction of obscured AGN being dependent on luminosity even
  if this quantity has no intrinsic luminosity
  dependence}. Specifically, in such a case, the observed obscured
fraction approaches a constant value in the limit of $L\ll\lb$, and
another, lower limiting value at $L\gg\lb$. It is easy to show that,
if the intrinsic ratio of obscured and unobscured AGN is
$(\nobsc/\nunobsc)_{\rm intr}$, then their observed ratio will be:
\beq
\left(\frac{\nobsc}{\nunobsc}\right)_{\rm obs}=
\left(\frac{\nobsc}{\nunobsc}\right)_{\rm intr}
\left(\frac{\langle\robsc\rangle}{\langle\runobsc\rangle}\right)^\gamma, 
\eeq
where $\langle\robsc\rangle$ and $\langle\runobsc\rangle$ are the
appropriately ensemble-averaged bias factors ($=\lobs/\lintr$) for
unobscured and  obscured AGN, respectively, and $\gamma$ is the
(effective) slope of the luminosity function. For example, in the case
of cosine-law emission and $\theta=30^\circ$, the corresponding bias
factors averaged over the log-uniform $\nheq$ distribution (see
Fig.~\ref{fig:flux_nh_theta}) are $\langle\runobsc\rangle\sim 2$ and
$\langle\robsc\rangle\sim 0.7$, whereas $(\nobsc/\nunobsc)_{\rm
  intr}=6.46$. Therefore, for our inferred intrinsic AGN LF, with
$\gamma\sim 0.9$ and $\gamma\sim 2.4$ in the low- and high-luminosity
ends, respectively, and $\lb\sim 10^{43.7}$, we may expect
$(\nobsc/\nunobsc)_{\rm obs}\sim 2.5$ and $(\nobsc/\nunobsc)_{\rm
  obs}\sim 0.5$ at $L\ll 10^{43.7}$ and $L\gg 10^{43.7}$~erg~s$^{-1}$,
respectively. This corresponds to the obscured AGN fractions
$\nobsc/(\nunobsc+\nobsc)\sim 0.71$ and $\sim 0.33$, respectively,
which is approximately what we see in Fig.~\ref{fig:toymodel_30} for
the results of simulations with the cosine-law emission law and
$\theta=30^\circ$. 

\begin{figure}
\centering
\includegraphics[width=0.95\columnwidth,bb=0 180 580 710]{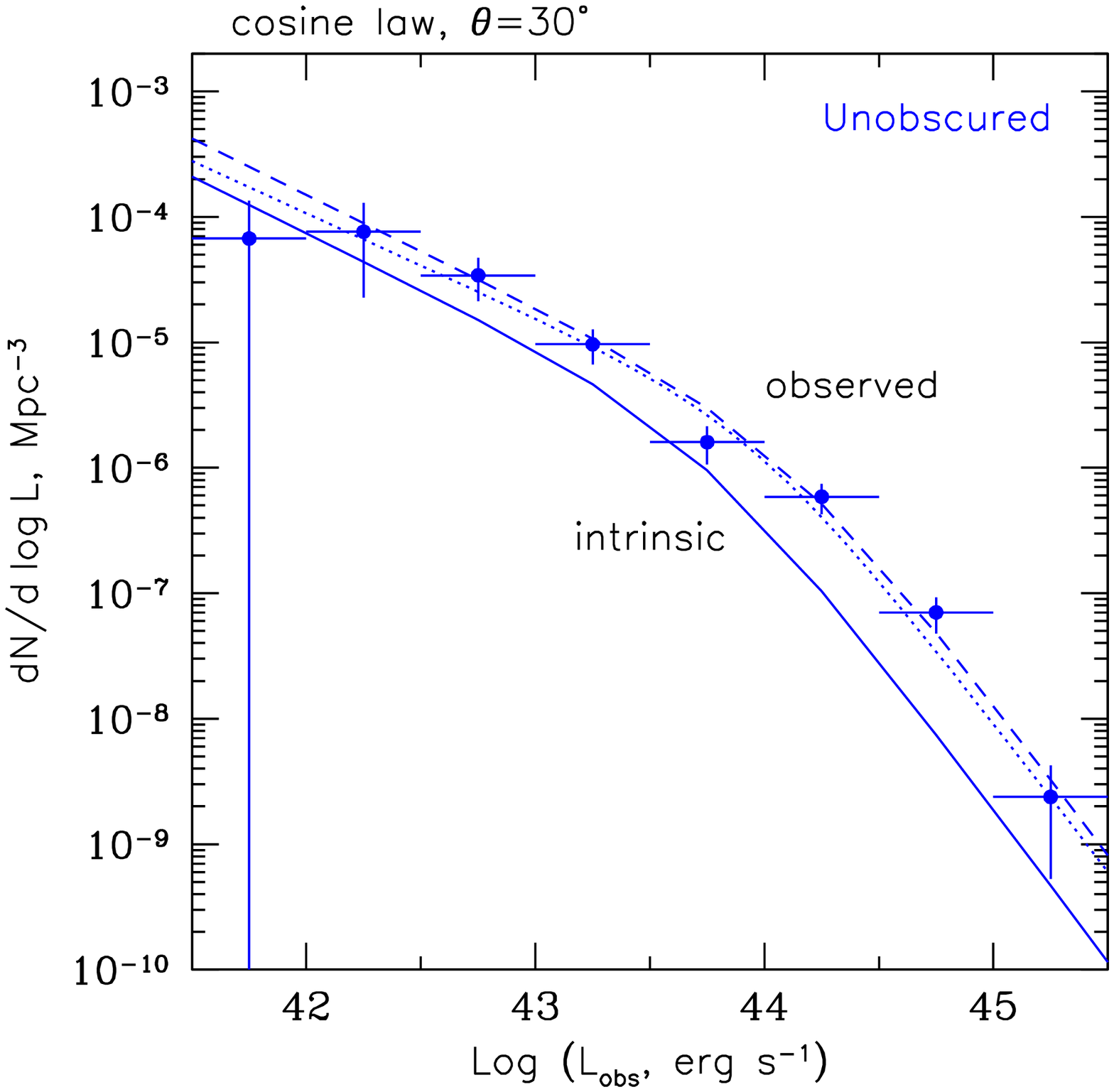}
\includegraphics[width=0.95\columnwidth,bb=0 180 580 710]{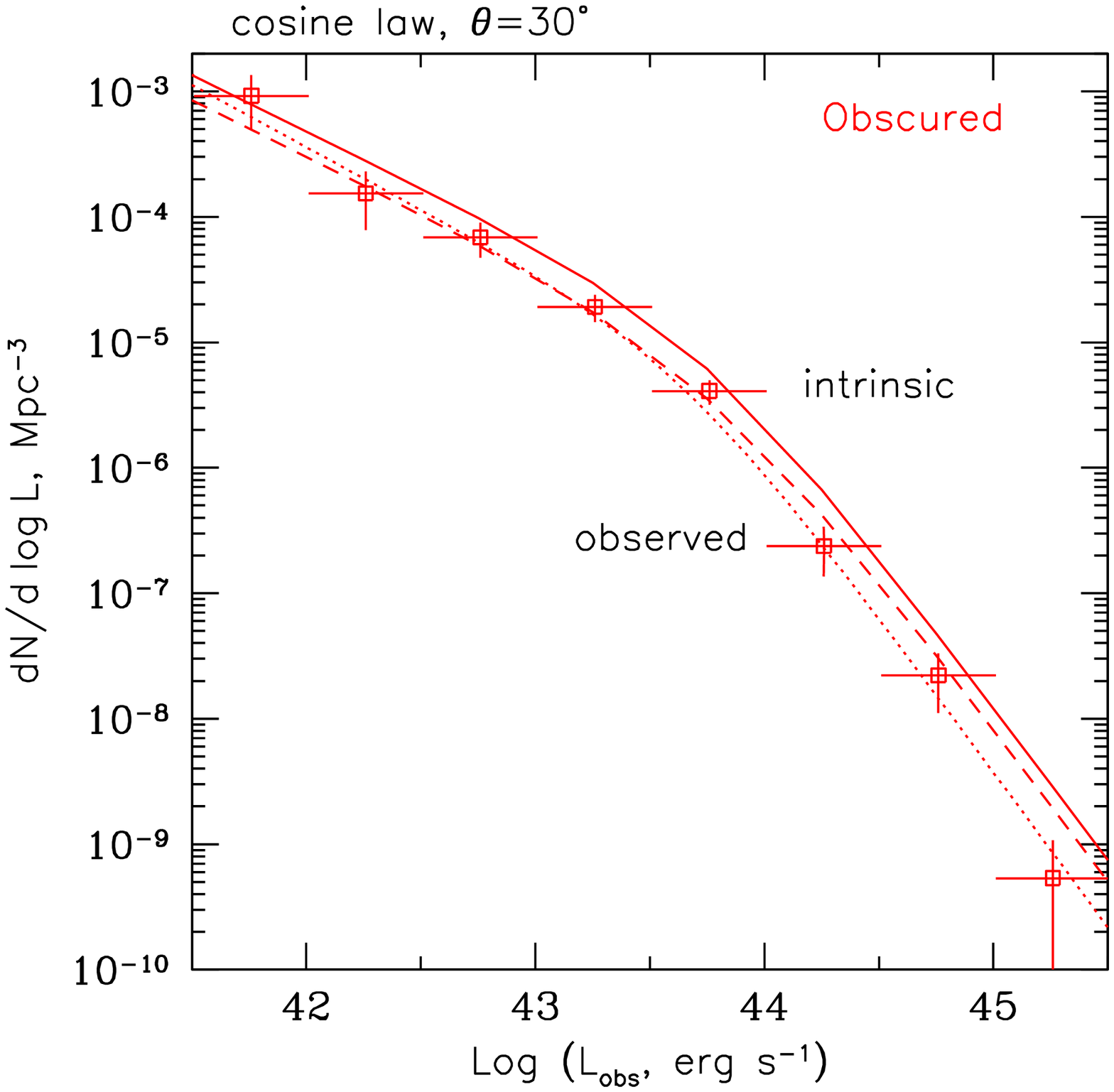}
\caption{\textit{Top:} Simulated observed LF of unobscured AGN (dashed
  line) for $\theta=30^\circ$, cosine-law emission, 
  $\nhmin=10^{22}$~cm$^{-2}$ and $\nhmax=10^{26}$~cm$^{-2}$, 
  in comparison with the LF of unobscured AGN observed by 
  \textit{INTEGRAL} (data points with error bars and their fit by a
  broken power-law model -- dotted line). The solid line shows the
  assumed intrinsic LF of unobscured AGN (equal to the total AGN LF
  multiplied by $(1-\cos\theta$)), which was previously (in
  \S\ref{s:intrinsic_lf}) derived from the \textit{INTEGRAL} data
  using an inverse approach. \textit{Bottom:} The same, but for
  obscured AGN. In this case the intrinsic LF of obscured AGN is
  equal to the total intrinsic LF multiplied by $\cos\theta$.
}
\label{fig:toymodel_lf}
\end{figure}

Figure~\ref{fig:toymodel_lf} shows how our direct
convolution model predicts the observed LFs of unobscured and
obscured AGN in the case of cosine-law emission and
$\theta=30^\circ$. We see that for unobscured AGN, the simulated 
observed LF fits the \textit{INTEGRAL} data well. In the 
case of obscured AGN, the match between the simulated and actually
observed LFs is good below $\lobs\sim 10^{44}$~erg~s$^{-1}$, but a
significant deviation is evident at higher luminosities. This probably
reflects the intrinsic differences between the inverse and direct
approaches to the considered problem, discussed at the beginning of this
section. Nevertheless, this difference does not significantly affect 
the conclusions of this study.

Note that in reality the situation may be more complicated. For
example, the intrinsic $\nheq$ distribution may depend on
luminosity. Also, the obscuring gas may be clumpy, so that the $\nh$
distribution observed for obscured AGN may represent not only the
distribution of torus column densities over the AGN population but
also the distribution of line-of-sight columns over different
observing directions for a given AGN torus. It would be interesting to
study this and other possibilities in future work, when significantly
larger samples of hard X-ray selected AGN become available.

\section{Discussion and summary}
\label{s:summary}

We utilised a sample of about 150 local ($z\lesssim 0.2$) hard X-ray
selected AGN, with reliable information on X-ray absorption columns,
to find out how strongly the observed declining trend of the obscured
AGN fraction with increasing luminosity may be affected by selection
effects. Using a torus-obscuration model and a state-of-the-art
radiative transfer code, we demonstrated that there must exist not
only a negative bias, due to absorption in the torus, in finding
obscured AGN in hard X-ray flux limited surveys, but also a positive
bias in detecting unobscured AGN -- due to reflection by the torus of
part of the radiation emitted by the central source towards the
observer. We further pointed out that these two biases may in fact
be even stronger if one takes into account plausible intrinsic
collimation of hard X-ray emission along the axis of the obscuring
torus, which can arise both in the hot corona where the hard X-ray
emission presumably originates and as a result of reflection of part
of this radiation by the underlying, optically thick accretion disk.

We demonstrated that for an AGN luminosity function that steepens at
high luminosities, which is indeed the case, these observational
biases should inevitably lead to the observed fraction of obscured AGN 
being smaller in the high-luminosity end of the LF than in the
low-luminosity end even if the obscured AGN fraction has no intrinsic
luminosity dependence. Moreover, even in the low-luminosity part of
the LF, the observed obscured fraction will be lower than its
intrinsic value. 

We explored two possibilities for the central hard X-ray source in
AGN: (i) isotropic emission and (ii) emission collimated 
according to Lambert's law, $dL/d\Omega\propto\cos\alpha$.  In the
former case, the intrinsic (i.e. corrected for the biases discussed
above) obscured AGN fraction reconstructed from our \textit{INTEGRAL}
sample still shows a declining trend with luminosity, although the
inferred intrinsic obscured fraction is larger than the observed one
at any luminosity. Namely, the obscured fraction is larger than $\sim
85$\% at $L\lesssim 10^{42.5}$~erg~s$^{-1}$ (17--60~keV), and 
decreases to $\lesssim 60$\% at $L\gtrsim 10^{44}$~erg~s$^{-1}$. In
terms of the half-opening angle $\theta$ of an obscuring torus, this
implies that $\theta\lesssim 30^\circ$ in lower-luminosity
AGN, and $\theta\gtrsim 45^\circ$ in higher-luminosity ones. If,
however, the emission from the central SMBH is collimated as
$dL/d\Omega\propto\cos\alpha$, then the derived intrinsic dependence
of the obscured AGN fraction is consistent with the opening angle of
the torus being constant with luminosity, namely $\theta\sim
30^\circ$.

At the moment, we regard both possiblities -- intrinsic obscuring AGN
fraction declining with luminosity or being constant -- as feasible,
as they depend on the presently poorly understood angular emission
diagram of the central source in AGN. We note however that a
luminosity-independent obscured AGN fraction might be consistent with
findings of some studies based on non-X-ray selected AGN samples (see
\citealt{lawelv10} for a discussion). We also note that the intrinsic
ratio of obscured to unobscured AGN that follows from our study, which
changes with luminosity from $\gtrsim 6:1$ to $\sim 1:1$ in the case
of isotropic emission and is $\sim 6:1$ in the case of cosine-law
emission, is not very different from the $\sim 4:1$ ratio inferred for
optically selected AGN by \citet{mairie95}. A more careful comparison
of these and other existing estimates of the ratio of obscured and
unobscured AGN in future work may help us get insight into the
geometrical and physical properties of obscuration in AGN, which may
be different in X-ray, optical, infrared and radio bands. 

The intrinsic dependence of the obscured AGN fraction on luminosity
derived here can find application in modelling the cosmic X-ray
background (CXB). Importantly, the inferred obscured fractions are
somewhat larger (even without allowance for possible intrinsic
collimation of X-ray emission in AGN) than those adopted in some
popular CXB synthesis models
(e.g. \citealt{treurr05,giletal07,uedetal14}) -- for example compare
our Figs.~\ref{fig:intr_obsc_frac_30} and \ref{fig:intr_obsc_frac_45}
with Fig.~13 in \citealt{giletal07} and Fig.~5 in \citealt{uedetal14}
(note, however, that these plots use the 2--10~keV energy band while
we use 17--60~keV).

As a byproduct, we reconstructed the intrinsic hard X-ray luminosity
function of local AGN and estimated the total number density and
luminosity density of AGN with $L>10^{40.5}$~erg~s$^{-1}$ (17--60~keV
and 2--10~keV), which may be used as reference $z=0$ values in the
study of cosmic AGN evolution and in modelling the cosmic X-ray
background.

The constraints on the intrinsic dependence of the obscured AGN
fraction on luminosity obtained in this work can be improved in the
near future using larger samples of hard X-ray selected AGN from
\textit{INTEGRAL}, \textit{Swift} and \textit{NuSTAR} surveys
(see \citealt{lanetal15} for a new constraint on the abundance
of heavily obscured AGN from \textit{NuSTAR} data). Note however that
it will be practially impossible to improve the current, fairly
uncertain estimate of the obscured AGN fraction at the highest
luminosities ($\gtrsim 10^{45}$~erg~s$^{-1}$) in the local Universe,
since the \textit{INTEGRAL} and \textit{Swift} all-sky surveys are
sensitive enough to detect all such objects in the local ($z\lesssim
0.2$) Universe and have found just a few of them because of the
very low space density thereof.

We finally note that AGN selection effects similar to those discussed
for hard X-ray surveys in this work should also affect obscured AGN
fractions inferred from samples selected in X-rays (at energies
below 10~keV). This should be studied in future work.  

\section*{Acknowledgments}

This study was supported by the Russian Foundation for Basic Research
(grant 13-02-01365). We thank the referee for useful comments.

\begin{appendix}

%

\section{AGN catalogue}
\label{s:catalog}

Table~\ref{tab:agn} presents the sample of non-blazar AGN at
$|b|>5^\circ$ used in this work. It is based on the catalogue of sources
\citep{krietal10b} detected during the \textit{INTEGRAL}/IBIS 7-year
all-sky survey. After publication of this catalogue, three previously
unidentified sources, IGR~J13466+1921, IGR~J14488$-$4009 and
IGR~J17036+3734 have been proved to be AGN and hence added to our
sample. For 31 nearby (closer than 40~Mpc) objects we
adopt distance estimates from the Extragalactic Distance Database
(EDD)\footnote{http://edd.ifa.hawaii.edu/}, whereas the distances of
the remaining objects are estimated from their redshifts, which are
adopted from NED. The quoted hard X-ray luminosities are observed
ones, calculated from the adopted distances and measured hard X-ray
(17--60~keV) fluxes \citep{krietal10b}.    

For this study, the most important AGN property is X-ray absorption
column density. The corresponding information has been updated with
respect to our previous publications
\citep{sazrev04,sazetal07,sazetal12} whenever necessary and
possible. The last column of Table~\ref{tab:agn} provides relevant
references. 

\begin{table*}
\caption{\textit{INTEGRAL} 7-year sample of non-blazar AGN at
  $|b|>5^\circ$.  
  \label{tab:agn}
} 

\begin{tabular}{lcrccrr}
\hline
\multicolumn{1}{c}{Object} &
\multicolumn{1}{c}{z} &
\multicolumn{1}{c}{$D$} & 
\multicolumn{1}{c}{Ref.} & 
\multicolumn{1}{c}{$\log{L_{17-60~{\rm keV}}}$} & 
\multicolumn{1}{c}{$\nh$} & 
\multicolumn{1}{c}{Ref.} \\

 & & \multicolumn{1}{c}{Mpc} & &
\multicolumn{1}{c}{erg~s$^{-1}$} &
\multicolumn{1}{c}{$10^{22}$~cm$^{-2}$} & \\
\hline
IGR J00040+7020    & 0.0960 &  442.4 & &  44.31 &     3 & 2\\
SWIFT J0025.8+6818 & 0.0120 &   52.0 & &  42.51 &$>1000$& 3\\
MRK 348            & 0.0150 &   65.2 & &  43.69 &    30 & 4\\
ESO 297-G018       & 0.0252 &  110.4 & &  43.86 &    50 & 5\\
IGR J01528$-$0326  & 0.0172 &   74.9 & &  43.06 &    14 & 6\\
NGC 788            & 0.0136 &   59.0 & &  43.33 &    40 & 4\\
MRK 1018           & 0.0424 &  188.1 & &  43.69 &  $<1$ & 7,8\\
IGR J02086$-$1742  & 0.1290 &  607.6 & &  44.77 &  $<1$ & 9\\
LEDA 138501        & 0.0492 &  219.4 & &  44.18 &  $<1$ & 10\\
MRK 590            & 0.0264 &  115.7 & &  43.26 &  $<1$ & 11\\
SWIFT J0216.3+5128 & 0.4220 & 2322.8 & &  46.15 &     3 & 7\\
MRK 1040           & 0.0167 &   72.7 & &  43.40 &  $<1$ & 12\\
IGR J02343+3229    & 0.0162 &   70.5 & &  43.26 &     2 & 13\\
NGC 985            & 0.0431 &  191.3 & &  43.91 &  $<1$ & 14\\
NGC 1052           & 0.0050 &   19.4 &1&  41.89 &    20 & 15\\
NGC 1068           & 0.0038 &   12.3 &1&  41.58 &$>1000$& 16\\
IGR J02524$-$0829  & 0.0168 &   73.1 & &  42.97 &    12 & 9\\
NGC 1142           & 0.0288 &  126.5 & &  43.99 &    50 & 4\\
NGC 1194           & 0.0136 &   59.0 & &  42.82 &$\sim100$?& 17\\
IGR J03249+4041    & 0.0476 &  212.0 & &  43.84 &     3 & 18\\
IGR J03334+3718    & 0.0550 &  246.3 & &  44.17 &  $<1$ & 7,18\\
NGC 1365           & 0.0055 &   18.0 &1&  42.16 &$\sim50$& 19\\
ESO 548-G081       & 0.0145 &   63.0 & &  43.21 &  $<1$ & 20\\
3C 105             & 0.0890 &  408.2 & &  44.69 &    30 & 7,8\\
3C 111             & 0.0485 &  216.1 & &  44.53 &  $<1$ & 21\\
IRAS 04210+0400    & 0.0450 &  200.0 & &  43.98 &    30 & 7,18\\
3C 120             & 0.0330 &  145.4 & &  44.14 &  $<1$ & 11\\
UGC 03142          & 0.0217 &   94.8 & &  43.59 &     3 & 7\\
CGCG 420-015       & 0.0294 &  129.2 & &  43.51 &$>1000$& 3\\
ESO 033-G002       & 0.0181 &   78.9 & &  43.05 &     1 & 22\\
LEDA 075258        & 0.0160 &   69.6 & &  42.75 &  $<1$ & 7,18\\
XSS J05054$-$2348  & 0.0350 &  154.4 & &  44.13 &     6 & 23\\
IRAS 05078+1626    & 0.0179 &   78.0 & &  43.66 &  $<1$ & 24\\
ARK 120            & 0.0327 &  144.0 & &  44.02 &  $<1$ & 25\\
ESO 362-G018       & 0.0124 &   53.8 & &  43.10 &  $<1$ & 26\\
PIC A              & 0.0351 &  154.9 & &  44.01 &  $<1$ & 11\\
NGC 2110           & 0.0078 &   29.0 &1&  43.11 &    14 & 27\\
MCG 8-11-11        & 0.0204 &   89.0 & &  43.92 &  $<1$ & 11\\
MRK 3              & 0.0135 &   58.6 & &  43.45 &   100 & 28\\
PMN J0623$-$6436   & 0.1289 &  607.1 & &  44.73 &  $<1$ & 29\\
IGR J06239$-$6052  & 0.0405 &  179.4 & &  43.80 &    20 & 30\\
IGR J06415+3251    & 0.0172 &   74.9 & &  43.39 &    16 & 31\\
MRK 6              & 0.0188 &   81.9 & &  43.52 &$\sim5$& 32\\
IGR J07563$-$4137  & 0.0210 &   91.7 & &  43.08 &  $<1$ & 33\\
ESO 209-G012       & 0.0405 &  179.4 & &  43.80 &  $<1$ & 34\\
IGR J08557+6420    & 0.0370 &  163.5 & &  43.64 &    20 & 7,18\\
IRAS 09149$-$6206  & 0.0573 &  257.0 & &  44.19 &  $<1$ & 10\\
IGR J09253+6929    & 0.0390 &  172.6 & &  43.66 &$>100$?& 35\\
NGC 2992           & 0.0077 &   29.0 &1&  42.75 &     1 & 36\\
MCG -5-23-16       & 0.0085 &   36.8 & &  43.27 &     2 & 11\\
IGR J09522$-$6231  & 0.2520 & 1276.8 & &  45.37 &     6 & 37\\
NGC 3081           & 0.0080 &   28.6 &1&  42.64 &   100 & 38\\
ESO 263-G013       & 0.0333 &  146.7 & &  43.72 &    30 & 39\\
NGC 3227           & 0.0039 &   26.4 &1&  42.84 &  $<1$ & 4\\
NGC 3281           & 0.0107 &   46.3 & &  43.08 &   200 & 40\\
IGR J10386$-$4947  & 0.0600 &  269.6 & &  44.05 &     1 & 4\\
IGR J10404$-$4625  & 0.0239 &  104.6 & &  43.48 &     3 & 4\\
NGC 3516           & 0.0088 &   38.0 &1&  42.87 &  $<1$ & 11\\
NGC 3783           & 0.0097 &   25.1 &1&  42.95 &  $<1$ & 4\\
IGR J11459$-$6955  & 0.2440 & 1230.9 & &  45.31 &  $<1$ & 7,41\\
IGR J12009+0648    & 0.0360 &  159.0 & &  43.67 &    11 & 31\\

\end{tabular}

\end{table*} 

\setcounter{table}{0}
\begin{table*}
\caption{(continued)  
} 

\begin{tabular}{lcrccrr}
\hline
\multicolumn{1}{c}{Object} &
\multicolumn{1}{c}{z} &
\multicolumn{1}{c}{$D$} & 
\multicolumn{1}{c}{Ref.} & 
\multicolumn{1}{c}{$\log{L_{17-60~{\rm keV}}}$} & 
\multicolumn{1}{c}{$\nh$} & 
\multicolumn{1}{c}{Ref.} \\

 & & \multicolumn{1}{c}{Mpc} & &
\multicolumn{1}{c}{erg~s$^{-1}$} &
\multicolumn{1}{c}{$10^{22}$~cm$^{-2}$} & \\
\hline
IGR J12026$-$5349  & 0.0280 &  122.9 & &  43.75 &     2 & 33\\
NGC 4051           & 0.0024 &   17.1 &1&  41.97 &  $<1$ & 11\\
NGC 4138           & 0.0030 &   13.8 &1&  41.60 &     8 & 42\\
NGC 4151           & 0.0033 &   11.2 &1&  42.66 &     8 & 11\\
IGR J12107+3822    & 0.0229 &  100.1 & &  43.14 &     3 & 7\\
NGC 4235           & 0.0080 &   31.5 &1&  42.06 &  $<1$ & 7\\
NGC 4253           & 0.0129 &   56.0 & &  42.74 &  $<1$ & 43\\
NGC 4258           & 0.0015 &    7.6 &1&  40.95 &     7 & 44\\
MRK 50             & 0.0234 &  102.4 & &  43.21 &  $<1$ & 24\\
NGC 4388           & 0.0084 &   16.8 &1&  42.78 &    40 & 36,45\\
NGC 4395           & 0.0011 &    4.7 &1&  40.55 &     2 & 46\\
NGC 4507           & 0.0118 &   51.2 & &  43.56 &$\sim70$& 47\\
ESO 506-G027       & 0.0250 &  109.5 & &  43.76 &   100 & 48\\
XSS J12389$-$1614  & 0.0367 &  162.1 & &  43.98 &     2 & 33\\
NGC 4593           & 0.0090 &   37.3 &1&  42.96 &  $<1$ & 4\\
WKK 1263           & 0.0244 &  106.8 & &  43.39 &  $<1$ & 31\\
NGC 4939           & 0.0104 &   34.7 &1&  42.36 &$>1000$?& 49\\
NGC 4945           & 0.0019 &    3.4 &1&  41.41 &   400 & 50\\
ESO 323-G077       & 0.0150 &   65.2 & &  43.06 &    30 & 4\\
IGR J13091+1137    & 0.0251 &  109.9 & &  43.64 &    60 & 39\\
IGR J13109$-$5552  & 0.1040 &  481.8 & &  44.71 &  $<1$ & 7\\
IGR J13149+4422    & 0.0366 &  161.7 & &  43.68 &     5 & 13\\
MCG -03-34-064     & 0.0165 &   71.8 & &  43.19 &$\sim50$& 36\\
CEN A              & 0.0018 &    3.6 &1&  41.99 &    11 & 27\\
ESO 383-G018       & 0.0124 &   53.8 & &  42.74 &    20 & 7,51\\
MCG -6-30-15       & 0.0077 &   25.5 &1&  42.48 &  $<1$ & 12\\
NGC 5252           & 0.0230 &  100.6 & &  43.91 &     5 & 36\\
MRK 268            & 0.0399 &  176.7 & &  43.81 &    30 & 7,18\\
IGR J13466+1921    & 0.0850 &  388.7 & &  44.62 &  $<1$ & 52\\
IC 4329A           & 0.0160 &   69.6 & &  44.02 &  $<1$ & 12\\
LEDA 49418         & 0.0509 &  227.2 & &  43.75 &     2 & 53\\
NGC 5506           & 0.0062 &   21.7 &1&  42.92 &     3 & 11\\
IGR J14175$-$4641  & 0.0766 &  348.3 & &  44.21 &$>100$?& 35\\
NGC 5548           & 0.0172 &   74.9 & &  43.27 &  $<1$ & 11\\
ESO 511-G030       & 0.0224 &   97.9 & &  43.45 &  $<1$ & 11\\
NGC 5643           & 0.0040 &   11.8 &1&  41.23 &$>1000$& 3\\
NGC 5728           & 0.0094 &   24.8 &1&  42.51 &   200 & 3\\
IGR J14488$-$4009  & 0.1230 &  577.1 & &  44.69 &     6 & 54\\
IGR J14552$-$5133  & 0.0160 &   69.6 & &  42.87 &  $<1$ & 10\\
IGR J14561$-$3738  & 0.0246 &  107.7 & &  43.20 &$\sim100$& 37\\
IC 4518A           & 0.0157 &   68.3 & &  43.07 &    10 & 7,18\\
MRK 841            & 0.0364 &  160.8 & &  43.98 &  $<1$ & 55\\
NGC 5995           & 0.0252 &  110.4 & &  43.66 &  $<1$ & 11\\
IGR J15539$-$6142  & 0.0149 &   64.8 & &  42.60 &    20 & 10\\
ESO 389-G002       & 0.0194 &   84.6 & &  42.93 &     6 & 18\\
WKK 6092           & 0.0156 &   67.8 & &  42.96 &  $<1$ & 24\\
IGR J16185$-$5928  & 0.0350 &  154.4 & &  43.55 &    10 & 56\\
ESO 137-G034       & 0.0092 &   33.0 &1&  42.26 &$>1000$& 39\\
IGR J16385$-$2057  & 0.0264 &  115.7 & &  43.27 &  $<1$ & 13\\
IGR J16482$-$3036  & 0.0313 &  137.7 & &  43.85 &  $<1$ & 4\\
NGC 6221           & 0.0050 &   15.6 &1&  41.66 &     1 & 57\\
NGC 6240           & 0.0245 &  107.3 & &  43.76 &   250 & 3\\
IGR J16558$-$5203  & 0.0540 &  241.6 & &  44.22 &  $<1$ & 4\\
IGR J17009+3559    & 0.1130 &  526.7 & &  44.74 &    30 & 7,18\\
IGR J17036+3734    & 0.0650 &  293.1 & &  44.36 &  $<1$ & 18\\
NGC 6300           & 0.0037 &   13.1 &1&  42.02 &    25 & 11\\
IGR J17418$-$1212  & 0.0372 &  164.4 & &  43.86 &  $<1$ & 4\\
H 1821+643         & 0.2970 & 1541.0 & &  45.58 &  $<1$ & 11\\
IC 4709            & 0.0169 &   73.6 & &  43.36 &    12 & 23\\
IGR J18249$-$3243  & 0.3550 & 1895.4 & &  45.61 &  $<1$ & 58\\
ESO 103-G035       & 0.0133 &   57.7 & &  43.43 &    30 & 11\\

\end{tabular}

\end{table*} 

\setcounter{table}{0}
\begin{table*}
\caption{(continued)  
} 

\begin{tabular}{lcrccrr}
\hline
\multicolumn{1}{c}{Object} &
\multicolumn{1}{c}{z} &
\multicolumn{1}{c}{$D$} & 
\multicolumn{1}{c}{Ref.} & 
\multicolumn{1}{c}{$\log{L_{17-60~{\rm keV}}}$} & 
\multicolumn{1}{c}{$\nh$} & 
\multicolumn{1}{c}{Ref.} \\

 & & \multicolumn{1}{c}{Mpc} & &
\multicolumn{1}{c}{erg~s$^{-1}$} &
\multicolumn{1}{c}{$10^{22}$~cm$^{-2}$} & \\
\hline
3C 390.3           & 0.0561 &  251.4 & &  44.59 &  $<1$ & 4\\
ESO 140-G043       & 0.0142 &   61.7 & &  43.14 &     2 & 59\\
ESO 025-G002       & 0.0289 &  126.9 & &  43.50 &  $<1$ & 7,18\\
IGR J18559+1535    & 0.0838 &  382.9 & &  44.54 &  $<1$ & 4\\
EXSS 1849.4$-$7831 & 0.0420 &  186.3 & &  44.02 &  $<1$ & 18\\
IGR J19077$-$3925  & 0.0760 &  345.4 & &  44.21 &  $<1$ & 7,18\\
IGR J19194$-$2956  & 0.1668 &  804.6 & &  45.03 &  $<1$ & 7,18\\
ESO 141-G055       & 0.0371 &  164.0 & &  44.06 &  $<1$ & 11\\
SWIFT J1930.5+3414 & 0.0633 &  285.1 & &  44.10 &    30 & 6\\
1H 1934$-$063      & 0.0106 &   45.9 & &  42.56 &  $<1$ & 60\\
IGR J19405$-$3016  & 0.0520 &  232.3 & &  43.98 &  $<1$ & 61\\
NGC 6814           & 0.0052 &   22.0 &1&  42.42 &  $<1$ & 12\\
XSS J19459+4508    & 0.0539 &  241.1 & &  44.03 &    11 & 33\\
CYG A              & 0.0561 &  251.4 & &  44.67 &    20 & 62\\
ESO 399-IG020      & 0.0250 &  109.5 & &  43.25 &  $<1$ & 7,56\\
IGR J20286+2544    & 0.0142 &   61.7 & &  43.21 &    50 & 2,4\\
4C +74.26          & 0.1040 &  481.8 & &  44.99 &  $<1$ & 11\\
MRK 509            & 0.0344 &  151.7 & &  44.21 &  $<1$ & 11\\
RX J2044.0+2833    & 0.0500 &  223.1 & &  44.08 &  $<1$ & 7,18\\
S5 2116+81         & 0.0860 &  393.6 & &  44.72 &  $<1$ & 63\\
IGR J21196+3333    & 0.0510 &  227.7 & &  43.95 &  $<1$ & 7,18\\
NGC 7172           & 0.0087 &   31.9 &1&  42.88 &    13 & 11\\
IGR J22292+6646    & 0.1120 &  521.7 & &  44.54 &  $<1$ & 58\\
NGC 7314           & 0.0048 &   15.9 &1&  41.92 &     1 & 64\\
MRK 915            & 0.0241 &  105.5 & &  43.45 &     3 & 7,18\\
MR 2251$-$178      & 0.0640 &  288.4 & &  44.72 &  $<1$ & 65\\
NGC 7465           & 0.0066 &   26.5 &1&  42.32 &$\sim10$& 66\\
NGC 7469           & 0.0163 &   70.9 & &  43.44 &  $<1$ & 4\\
MRK 926            & 0.0469 &  208.8 & &  44.28 &  $<1$ & 4\\
\hline
\end{tabular}

\textbf{References:} (1) Distance adopted from the Extragalactic
Distance Database, rather than calculated from the redshift; (2)
\cite{deretal12}; (3) \cite{krietal15}; (4) \cite{sazetal07}; (5)
\cite{uedetal07}; (6) \cite{lanetal07}; (7) \cite{sazetal10}; (8)
\cite{winetal09a}; (9) \cite{rodetal10}; (10) \cite{maletal07}; (11)
\cite{sazrev04}; (12) \cite{reyetal97}; (13) \cite{rodetal08}; (14)
\cite{kroetal09}; (15) \cite{teretal02}; (16) \cite{bauetal14}; (17)
\cite{greetal08}; (18) \cite{maletal12}; (19) \cite{waletal14}; (20)
\cite{paretal09}; (21) \cite{lewetal05}; (22) \cite{vigetal98}; (23)
\cite{revetal06}; (24) \cite{moletal09}; (25) \cite{vauetal04}; (26)
\cite{waletal13}; (27) \cite{fuketal11}; (28) \cite{ikeetal09}; (29)
\cite{galetal06}; (30) \cite{revetal07}; (31) \cite{winetal08}; (32)
\cite{immetal03}; (33) \cite{sazetal05}; (34) \cite{panetal08}; (35)
this work; (36) \cite{risaliti02}; (37) \cite{sazetal08}; (38)
\cite{eguetal11}; (39) \cite{cometal10}; (40) \cite{vigcom02}; (41)
\cite{lanetal10}; (42) \cite{capetal06}; (43) \cite{turetal07}; (44)
\cite{youwil04}; (45) \cite{shietal08}; (46) \cite{moretal05}; (47)
\cite{braetal13}; (48) \cite{winetal09b}; (49) \cite{maietal98},
strongly variable $\nh$; (50) \cite{pucetal14}; (51) \cite{nogetal09}; (52)
\cite{vasetal13}; (53) \cite{risetal00}; (54) \cite{moletal12}; (55)
\cite{petetal07}; (56) \cite{panetal11}; (57) \cite{levetal01}; (58)
\cite{lanetal09}; (59) \cite{ricetal10}; (60) \cite{maletal08}; (61)
\cite{zhaetal09}; (62) \cite{youetal02}; (63) \cite{moletal08}; (64)
\cite{dewgri05}; (65) \cite{reetur00}; (66) \cite{guaetal05b}.

\end{table*}

\end{appendix}


\end{document}